\documentclass[fleqn,usenatbib]{mnras}

\usepackage{newtxtext,newtxmath}
\usepackage[T1]{fontenc}

\DeclareRobustCommand{\VAN}[3]{#2}
\let\VANthebibliography\thebibliography
\def\thebibliography{\DeclareRobustCommand{\VAN}[3]{##3}\VANthebibliography}

\usepackage{graphicx}	% Including figure files
\usepackage{amsmath}	% Advanced maths commands
\def \pks  {{PKS~1830$-$211}} 
\newcommand\kms{\ifmmode {\rm km\ s}^{-1} \else km s$^{-1}$\fi} 
\newcommand\FWHM{\ifmmode {\rm FWHM} \else ${\rm FWHM}$\fi}
\newcommand\Lsun{\ifmmode L_{\odot} \else $L_{\odot}$\fi} 
\newcommand\Msun{\ifmmode M_{\odot} \else $M_{\odot}$\fi} 
\newcommand\Hbeta{\ifmmode {\rm H}\beta 
 \else H$\beta$\fi}

\def \ergsec{\hbox{erg s$^{-1}$}}
\def \ferg {erg cm$^{-2}$ s$^{-1}$}
\def \hcm {\hbox {\ifmmode $ atom cm$^{-2}\else atom cm$^{-2}$\fi}}

\def \arcsec {\hbox{$^{\prime\prime}$}}

\def \gray {$\gamma$-ray}
\def \phcmsec{\hbox{photons cm$^{-2}$ s$^{-1}$}}

\def \ergsc{\hbox{erg s$^{-1}$ cm$^{-2}$}}

\def \colc {cm$^{-2}$}
\def \grid{AGILE-{GRID}}

\newcommand{\asec} {\mbox{$^{\prime \prime}$} }
\newcommand{\amin} {\mbox{$^{\prime}$}}

\title[PKS 1830-211 2019 \gray{} flare observations]{Multi-wavelength observations of the lensed quasar PKS~1830$-$211 during the 2019 \gray{} flare}

\author[S. Vercellone et al.]{
S. Vercellone,$^{1}$\thanks{E-mail: stefano.vercellone@inaf.it (SV)}
I. Donnarumma$^{2}$, 
C. Pittori$^{3, 4}$, 
F. Capitanio$^{5}$, 
A. De Rosa$^{5}$, 
L. Di Gesu$^{2}$, 
\newauthor S. Kiehlmann$^{6, 7}$, 
M. N. Iacolina$^{2}$, 
P. A. Pellizzoni$^{8}$, 
E. Egron$^{8}$, 
L. Pacciani$^{5}$, 
G. Piano$^{5}$, 
S. Puccetti$^{2}$, 
\newauthor S. Righini$^{9}$, 
G. Valente$^{2}$, 
F. Verrecchia$^{3, 4}$, 
V. Vittorini$^{5}$, 
M. Tavani$^{5}$, 
E. Brocato$^{3, 10}$, 
A. W. Chen$^{11}$, 
\newauthor T. Hovatta$^{12, 13}$, 
A. Melis$^{8}$,
W. Max--Moerbeck$^{14}$, 
D. Perrodin$^{8}$, 
M. Pilia$^{8}$, 
M. Pili$^{8}$, 
A. C. S. Readhead$^{15}$, 
\newauthor R. Reeves$^{16}$, 
A. Ridolfi$^{8}$, 
F. Vitali$^{3}$, 
A. Bulgarelli$^{17}$, 
P. W. Cattaneo$^{18}$, 
F. Lucarelli$^{3, 4}$, 
A. Morselli$^{19}$,
A. Trois$^{8}$ \\
$^{1}$INAF, Osservatorio Astronomico di Brera, Via Emilio Bianchi 46, I-23807 Merate (LC), Italy\\
$^{2}$ASI, Via del Politecnico, I-00133 Roma, Italy\\
$^{3}$INAF, Osservatorio Astronomico di Roma, via Frascati 33, I-00078 Monte Porzio Catone, Italy\\
$^{4}$ASI Space Science Data Center, Via del Politecnico, I-00133 Roma, Italy\\
$^{5}$INAF, Istituto di Astrofisica e Planetologia Spaziali, Via Fosso del Cavaliere 100, I-00133 Roma, Italy\\
$^{6}$Institute of Astrophysics, Foundation for Research and Technology-Hellas, GR-71110 Heraklion, Greece\\
$^{7}$Department of Physics, Univ. of Crete, GR-70013 Heraklion, Greece\\
$^{8}$INAF, Osservatorio Astronomico di Cagliari, Via della Scienza 5, I-09047 Selargius (CA), Italy\\
$^{9}$INAF, Istituto di Radioastronomia, via Piero Gobetti 93/2, I-40129 Bologna, Italy\\
$^{10}$INAF, Osservatorio Astronomico d'Abruzzo, Via Mentore Maggini, I-64100 Teramo, Italy\\
$^{11}$School of Physics, University of the Witwatersrand, 1 Jan Smuts Avenue, Braamfontein 2000 Johannesburg, South Africa\\
$^{12}$Finnish Center for Astronomy with ESO (FINCA), University of Turku, FI-20014, Turku, Finland\\
$^{13}$Aalto University Mets\"{a}hovi Radio Observatory, Mets\"{a}hovintie 114, 02540 Kylm\"{a}l\"{a}, Finland\\
$^{14}$Departamento de Astronom{\'i}a, Universidad de Chile, Camino El Observatorio 1515, Las Condes, Santiago, Chile\\
$^{15}$Owens Valley Radio Observatory, California Institute of Technology, Pasadena, CA 91125, USA\\
$^{16}$Departamento de Astronom{\'i}a, Universidad de Concepci{\'o}n, Concepci{\'o}n, Chile\\
$^{17}$INAF, Osservatorio di Astrofisica e Scienza dello Spazio, via Gobetti 101, I-40129 Bologna (BO), Italy\\
$^{18}$INFN Sezione di Pavia, via U. Bassi 6, I-27100 Pavia (PV), Italy\\
$^{19}$INFN Sezione di Roma Tor Vergata, Via della Ricerca Scientifica 1, I-00133 Roma, Italy
}

\date{Accepted XXX. Received YYY; in original form ZZZ}

\pubyear{2023}

\begin{document}
\label{firstpage}
\pagerange{\pageref{firstpage}--\pageref{lastpage}}
\maketitle

\begin{abstract}
\pks{} is a \gray{} emitting, high-redshift (z $= 2.507 \pm 0.002$), lensed flat-spectrum radio quasar. During the period mid-February to mid-April 2019, this source underwent a series of strong \gray{} flares that were detected by both AGILE-GRID and {\it Fermi}-LAT, reaching a maximum \gray{} flux of $F_{\rm E>100\,MeV}\approx 2.3\times10^{-5}$\,\phcmsec. Here we report on a coordinated campaign from both on-ground (Medicina, OVRO, REM, SRT) and orbiting facilities (AGILE, {\it Fermi}, INTEGRAL, {\it NuSTAR}, {\it Swift}, {\it Chandra}), with the aim of investigating the multi-wavelength properties of \pks{} through nearly simultaneous observations presented here for the first time. We find a possible break in the radio spectra in different epochs above 15\,GHz, and a clear maximum of the 15\,GHz data approximately 110\,days after the \gray{} main activity periods. The spectral energy distribution shows a very pronounced Compton dominance (> 200) which challenges the canonical one-component emission model. Therefore we propose that the cooled electrons of the first component are re-accelerated to a second component by, e.g., kink or tearing instability during the \gray{} flaring periods. We also note that \pks{} could be a promising candidate for future observations with both Compton satellites (e.g., e-ASTROGAM) and Cherenkov arrays (CTAO) which will help, thanks to their improved sensitivity, in extending the data availability in energy bands currently uncovered.
\end{abstract}


%%%%%%%%%%%%%%%%%%%%%%%%%%%%%%%%%%%%%%%%%%%%%%%%%%

%%%%%%%%%%%%%%%%% BODY OF PAPER %%%%%%%%%%%%%%%%%%

%%%%%%%%%%%%%%%%%%%%%%%%%%%%%%%%
\section{Introduction}\label{introduction}
%%%%%%%%%%%%%%%%%%%%%%%%%%%%%%%%
{PKS~1830$-$211} is a high-redshift blazar~\citep[z $= 2.507 \pm 0.002$;][]{1999ApJ...514L..57L} that is gravitationally lensed by a spiral galaxy at z$ = 0.886$ \citep{1996Natur.379..139W}, as shown by the two radio lobes located 1" apart from each other \citep[A e B components,][]{1998ApJ...508L..51L}. The lensed counterparts were also observed in the near-IR (NIR) and optical energy bands by the Hubble Space Telescope and the Gemini Observatory~\citep{2002ApJ...575...95C}. The source was observed in X-rays by both XMM-Newton and {\it Chandra}, enabling for a study of the complex soft X-ray behavior in detail~\citep{2005A&A...438..121D, 2008AJ....135..333D}. 
\pks{} is a well-known \gray{} source above 100\,MeV, identified as such by~\citet{1997ApJ...481...95M} and subsequently listed in both the AGILE~\citep{2013A&A...558A.137V, 2019A&A...627A..13B} and {\it Fermi}-LAT~\citep{2020ApJS..247...33A} catalogues, which has produced several $\gamma$-ray flares over the last two decades~\citep{1999ApJS..123...79H, 2011ApJ...736L..30D, 2015ApJ...799..143A}.
A quite bright $\gamma$-ray flare was detected by AGILE and {\it Fermi}-LAT in October-November 2010, and multi-wavelength observations were carried out, as reported in \cite{2011ApJ...736L..30D} and \cite{2015ApJ...799..143A}. The multi-wavelength campaign carried out in 2010 with AGILE \citep{2011ApJ...736L..30D} showed that the intense $\gamma$-ray flare had no significant counterpart at lower frequencies, making this blazar classified as a ``$\gamma$-ray only flaring blazar''. This behaviour was discussed according to both (macro/micro) lensing and intrinsic physical properties of the blazar. In particular, macro and micro-lensing were excluded, given the chromaticity of the flare and the time scale of the $\gamma$-ray variability \citep{2011ApJ...736L..30D}. The flare was therefore associated with intrinsic variations of the jet emission, which are difficult to be interpreted in the one zone leptonic model~\citep{1985A&A...146..204G} given the high Compton dominance (i.e., the ratio of the peak of the Compton to the synchrotron peak luminosities) of the typical two-bump spectral energy distribution (SED) in flat-spectrum radio quasars. Alternative models have been invoked to overcome the limits of the one-zone leptonic models, such as the ``mirror model''~\citep{2015ApJ...814...51T, 2017ApJ...843L..23V} or the ``jet-cloud interaction model''~\citep{2010A&A...522A..97A,2014ApJ...793...98V}.
A long term $\gamma$-ray monitoring program represents an optimal tool to search for the time delay between the emissions of the two lensed images A and B as measured in the radio maps~\citep[26$^{+4}_{-5}$ days;][]{1998ApJ...508L..51L}. 

AGILE did not detect any delay between the lensed components during the $\gamma$-ray activity recorded in October--November 2010, which would infer a lack of delay if the flux ratio of the two components is $\sim 1$, as observed in radio (no conclusion could be drawn if this value was below 1 due to the AGILE sensitivity). The missing evidence of an echo can be explained with a flux ratio of the two components not equal to 1 in $\gamma$-rays. The dependence on energy of the flux ratio of the two components can be explained by invoking micro-lensing effects~\citep{2006ApJ...640..569B, 2011ApJ...736L..30D, 2015ApJ...799..143A}. Further analyses of the $\gamma$-ray light curve with a larger sensitivity were performed by~\cite{2015ApJ...799..143A}. By scanning a longer period (August 2008 - July 2011), two large $\gamma$-ray flares of \pks{} were detected by {\it Fermi}-LAT with no evidence for a delayed activity. Nevertheless,~\cite{2015ApJ...799..143A} were able to place a lower limit of $\sim$ 6 on the flux ratio between the two lensed images. \citet{2015ApJ...809..100B}, analysing {\it Fermi}-LAT data between August 2008 and January 2015 found a \gray{} time-delay consistent with the radio one, while~\citet{2021ApJ...915...26A}, analysing a time period in 2019 similar to the one discussed in this paper, found no clear evidence of such a \gray{} time-delay.

In this paper, we present the multi-frequency campaign on \pks{} during the period mid-February to mid-April 2019 and involving measurements in the radio, near infra-red, optical, UV, X-ray and $\gamma$-ray energy bands. 
The paper is organised as follows. Section~\ref{SEC:obs:summary:rev1} reports on the different facilities involved in this observing campaign. In Section~\ref{SEC:discussion} we discuss our results, while in Appendices~\ref{SEC:obs:gamma} --~\ref{SEC:obs:radio} we report the multi-wavelength observations.
We adopt a $\Lambda$-cold dark matter ($\Lambda$-CDM) cosmology~\citep{2020A&A...641A...6P} with the following parameters H$_{\rm 0}$ = 67.7, $\Omega_{m}$ = 0.31, and $\rm \Omega_{\Lambda}= 0.69$.

%%%%%%%%%%%%%%%%%%%%%%%%%%%%%%%%
\section{Summary of Observations and Flare Definition}\label{SEC:obs:summary:rev1}
%%%%%%%%%%%%%%%%%%%%%%%%%%%%%%%%
The detection of \gray{} flares from \pks{} triggered a large multi-wavelength observing campaign, involving both on-ground (Medicina, OVRO, REM, SRT) and orbiting facilities (AGILE, {\it Fermi}, INTEGRAL, {\it NuSTAR}, {\it Swift}, {\it Chandra}). %
%
%
%_____________________________________________________________
   \begin{figure}
   \centering
   \includegraphics[angle=0,width=9cm]{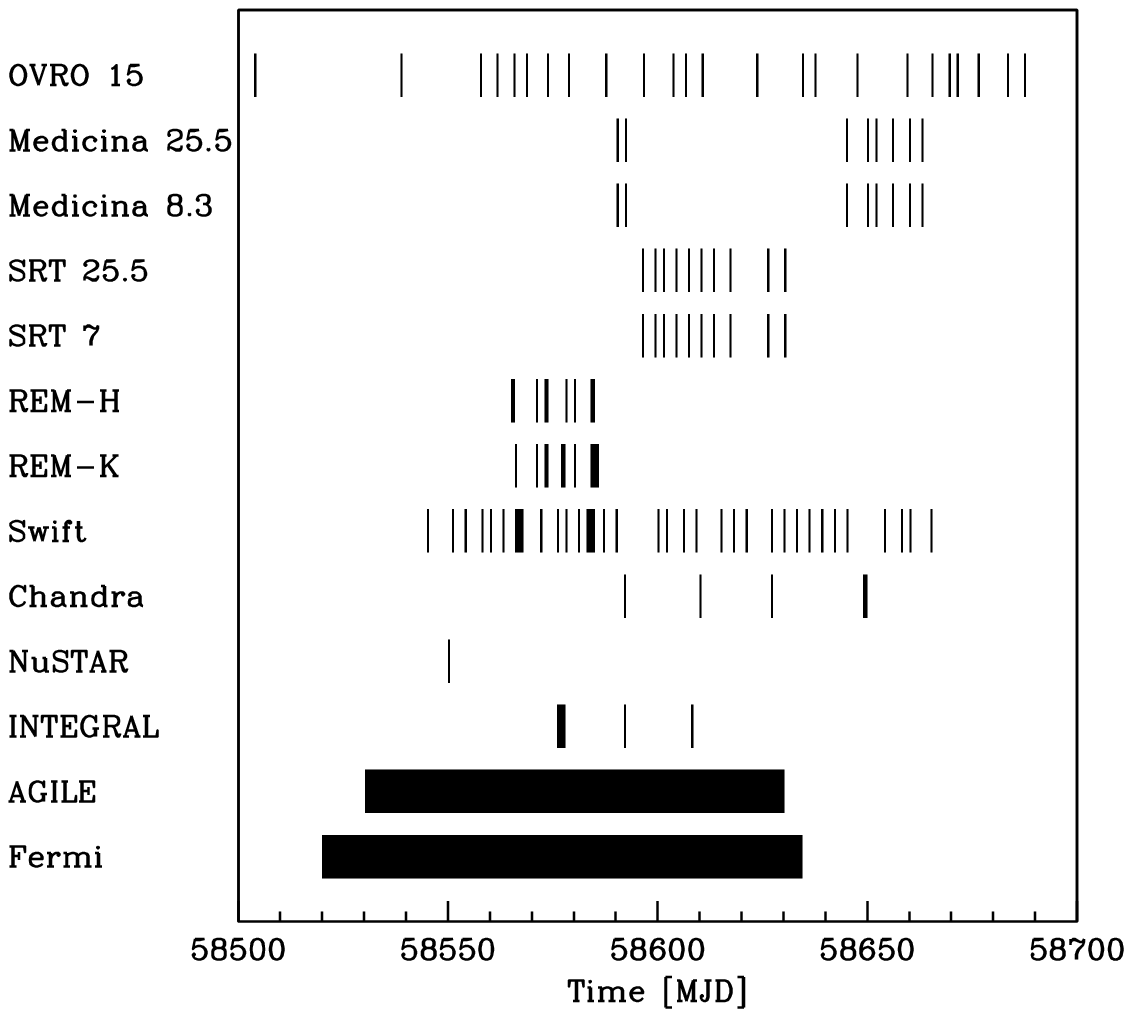}
      \caption{Coverage of the different facilities as a function of time. The first five values in the ordinate label refer to the observing frequency in GHz.}
         \label{Fig:ObservatoriesGantt}
   \end{figure}
%_____________________________________________________________
%
%

%
%_____________________________________________________________________
% OBSERVATORIES
%----------------------------------------------------------------------
\begin{table}
\caption{List of facilities and their energy range.}     
\label{TAB:Observatories}      
\centering    
\begin{tabular}{ll}        
\hline        
Facility   &  Energy/   \\
              &  Wavelength/   \\
              &  Frequency   \\
\hline
{\it Fermi}-LAT  & 0.1-300\,GeV  \\
AGILE-GRID  & 0.1-30\,GeV  \\
INTEGRAL-ISGRI  & 13-200\,keV  \\
{\it NuSTAR} & 3-78.4\,keV \\
{\it Chandra}-ACIS & 0.5-8\,keV \\
{\it Swift}-XRT & 0.3-10\,keV \\
{\it Swift}-UVOT & {\it v} 5468\,\AA \\
                          & {\it b} 4392\,\AA \\
                          & {\it u} 3465\,\AA \\
                          & {\it w1} 2600\,\AA \\
                          & {\it m2} 2246\,\AA \\
                          & {\it w2} 1928\,\AA \\
REM-REMIR     & {\it K} 22000\,\AA \\
REM                  & {\it H} 16350\,\AA \\
SRT                   & {\it C-Band} 7\,GHz \\
                         & {\it K-Band} 25.5\,GHz \\
Medicina           & {\it X-Band} 8.3\,GHz \\
                         & {\it K-Band} 25.5\,GHz \\
OVRO              & 15\,GHz\\
\hline
\end{tabular}
\end{table}
Figure~\ref{Fig:ObservatoriesGantt} shows the multi-wavelength coverage as a function of time of the different instruments, while Table~\ref{TAB:Observatories} reports the different energies covered by our campaign. These observations allow us to reconstruct an almost simultaneous spectral energy distribution spanning about fifteen decades in energy.
The detailed description of each facility, data reduction, data analysis and the presentation of the results can be found in Appendix~\ref{SEC:obs:gamma} (\gray{} data), Appendix~\ref{SEC:obs:Xrays} (X-ray data),  Appendix~\ref{SEC:obs:iropt} (IR, Optical, and UV data), and Appendix~\ref{SEC:obs:radio} (radio data).
The {\it Fermi}-LAT 12-hr binning allows us to obtain a detailed description of the different \gray{} flares. We anticipate here the method we used for selecting the different \gray{} flares time-intervals. 

\begin{figure}
   \centering
   \includegraphics[angle=0,width=9cm]{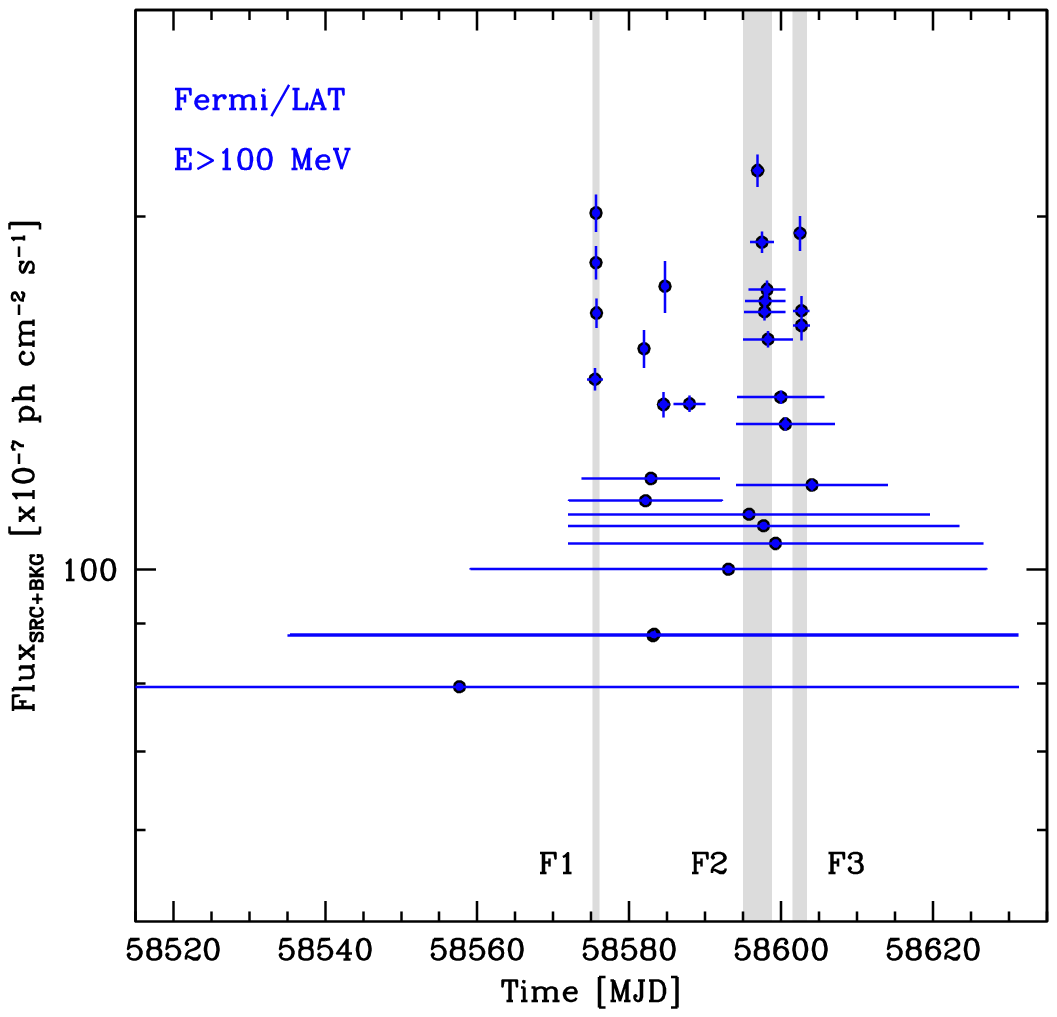}
      \caption{\pks{} {\it Fermi}-LAT unbinned light curve (a data clustering following \citealt{2018A&A...615A..56P}). The most significant flares (and their durations) are reported as dashed areas. Background contribution is not subtracted. Dashed vertical bands correspond, from left to right, to F1, F2 and F3, respectively.}
         \label{Fig:FLAT.CLUSTERS}
   \end{figure}

``Unbinned light curves'' were produced for the brightest flaring periods following the procedure described in \cite{2018A&A...615A..56P}, by means of a photometric method. Gamma-rays are collected within an extraction region of radius $R_{68\%}^i(E)$ that varies with the energy and type of the reconstructed $\gamma$-ray. $R_{68\%}^i(E)$ corresponds to the 68\% containment radius of $\gamma$-rays of energy $E$ and reconstruction topology PSF$_i$ (point-spread function, with $i=1,...,4$).   {A novel method to select flares within a set of time-tagged data has been used}. It is a clustering method (iSRS, iterated short range search) in one dimension (the cumulative exposure domain), followed by a statistical discrimination based on maximum-score scan statistics \citep{glaz2006}. Once the set of collected $\gamma$-rays are produced, a clustering scheme in the cumulative exposure  domain is performed. Maximum-score scan statistics is applied to remove statistically not-relevant clusters~\citep[see][for details]{2018A&A...615A..56P}. The threshold chance probability to discriminate non-relevant clusters is set to 1.3\textperthousand. The set of remaining clusters is a root, with leaves corresponding to the detected peaks. Each cluster can be described by its mean time, the average flux within the cluster, and its length in time domain. In Figure~\ref{Fig:FLAT.CLUSTERS} we show the set of remaining clusters (represented by a segment) for \pks{}. We call this set the unbinned light curve. The most significant flares are F1 (MJD~58575.2--58576.1), F2 (MJD~58595.0--58598.8), and F3 (MJD~58601.5--58603.4).
Table~\ref{TAB:FLAT_Spectra} shows the {\it Fermi}-LAT spectral properties during the different flaring periods, assuming a log-parabola model, 
\begin{equation}\label{eqn:LP}
\frac{dN}{dE}=K_0 \left( \frac{E}{E_0} \right) ^{-\alpha -\beta \ln \left(E/E_0\right)},
\end{equation}
where $\alpha$ is the spectral slope, $\beta$ the curvature. The last value refers to the 4FGL Catalog one.
%
%_____________________________________________________________________
% Fermi-LAT spectral parameters
%----------------------------------------------------------------------
\begin{table}
\caption{{\it Fermi}-LAT spectral properties during the different flaring periods.}     
\label{TAB:FLAT_Spectra}      
\centering 
\begin{tabular}{lccc}      
\hline        
Period &  F$_{\rm E>100\,MeV}$  &  $\alpha$  &  $\beta$  \\
           &  ($10^{-5}$\phcmsec)  &  &  \\
\hline
F1     &  $1.95 \pm 0.07$  &   $2.44 \pm 0.05$   &   $0.11 \pm 0.03$  \\
F2     &  $2.07 \pm 0.07$  &   $2.26 \pm 0.03$   &   $0.15 \pm 0.02$  \\
F3     &  $1.77 \pm 0.07$  &   $2.33 \pm 0.03$   &   $0.12 \pm 0.02$  \\
4FGL &  $(4.47 \pm 0.12) \times 10^{-2}$  &   $2.46 \pm 0.01$   &   $0.09 \pm 0.01$  \\
\hline
\end{tabular}
\end{table}
%
%
%

%
%_____________________________________________________________
%                                    One column rotated figure
%-------------------------------------------------------------
%

%
%
%
%

%%%%%%%%%%%%%%%%%%%%%%%%%%%%%%%%
\section{Results and Discussion}\label{SEC:discussion}
%%%%%%%%%%%%%%%%%%%%%%%%%%%%%%%%
%

%%%%%%%%%%%%%%%%%%%%%%%%%%%%%%%%
\subsection{Multi-wavelength data}\label{Disc:MWL}
\begin{figure*}
   \centering
   \includegraphics[angle=0,width=\textwidth]{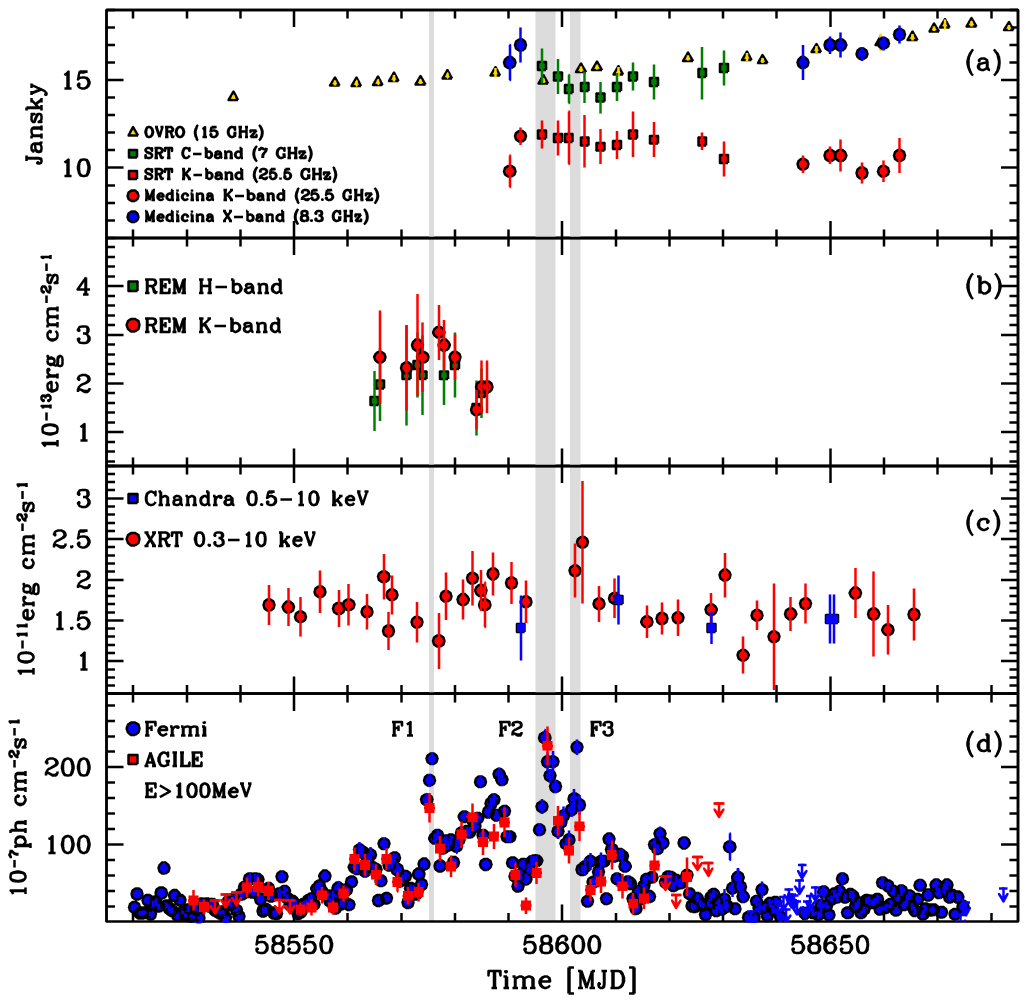}
   \vspace{-1 truecm}
      \caption{\pks{} multi-wavelength light-curves. From top to bottom: radio (7, 8.3, 15, 25.5)\,GHz, IR (H-band, K-band), X-ray (0.3--10)\,keV, and \gray{} (E>100\,MeV) data. The dashed areas correspond to the major \gray{} flares F1, F2 and F3, when the spectral energy distributions were computed. Arrows mark $3\sigma$ upper limits.}
         \label{Fig:MWL.LC}
   \end{figure*}

%%%%%%%%%%%%%%%%%%%%%%%%%%%%%%%%
%
Figure~\ref{Fig:MWL.LC} shows the \pks{} multi-wavelength light-curves. From top to bottom we show the radio (OVRO 15\,GHz, Medicina~8.3\,GHz \& 25.5\,GHz, SRT~7\,GHz \& 25.5\,GHz), IR (REM H-band and K-band), X-ray ({\it Swift}/XRT, {\it Chandra}), and \gray{} (AGILE-GRID and {\it Fermi}-LAT) data, respectively. 
The source reached its maximum flux ($F_{\rm E>100\,MeV} = (2.28\pm0.25)\times10^{-5}$\,\phcmsec) around April 24 (MJD$=58597.25\pm1.0$), as shown in Panel (d). This flux level is unprecedented for this source, and it is one of the largest ever detected in $\gamma$-rays from blazars at redshift $z > 2$~\citep[see][for 4C$+$71.07 at $z=2.172$]{2019A&A...621A..82V}.
Our {\it Chandra} data show that the flux is somewhat higher than what reported in~\citet{2005A&A...438..121D} for the {\it Chandra} observations of 2000-2001 and we also find an intervening column density N$_{\rm H,lens}$ slightly higher than what found in~\citet{2005A&A...438..121D}. To further investigate the possible variability of the spectral parameters of the source, we compared the count-rate observed in the two lensed images of PKS~1830$-$211 in a soft (0.5-2.0\,keV) and in a hard (2.0-8.0\,keV) energy band, as shown in Figure~\ref{Fig:CHANDRA_HOTSPOTS}. In the last two shorter observations, the ratio is unconstrained in the soft band. A slight decrease of the ratio N/S is observed, in agreement with that reported in~\citet{2019ATel12737....1W} for the ObsID 22197-22198. The decrease is more noticeable in the soft X-ray domain, which suggests variability of the absorbing column density in the lensing galaxy. However, higher quality, spatially resolved X-ray data would be needed to draw firm conclusions.
%
%_____________________________________________________________
%                                    One column rotated figure
%-------------------------------------------------------------

\begin{figure}
   \centering
   \includegraphics[angle=0,width=9cm]{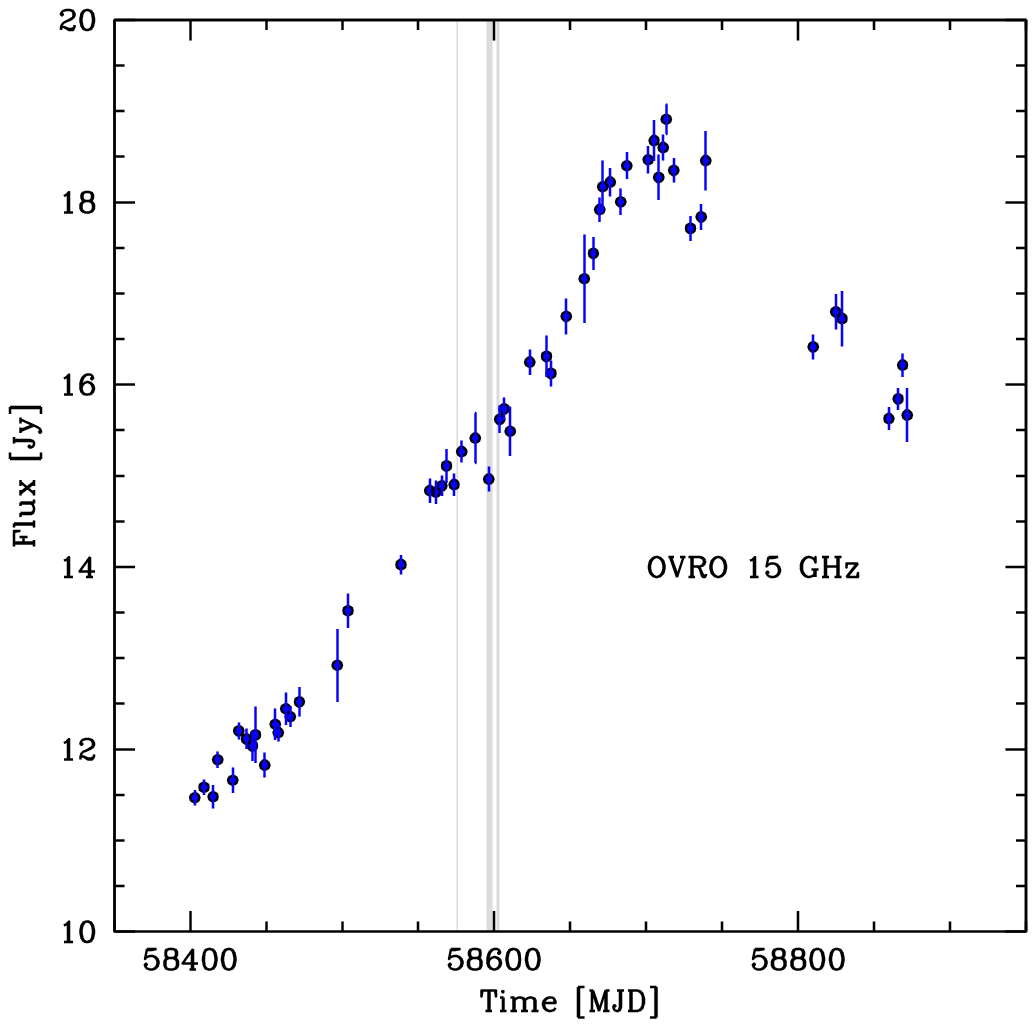}
      \caption{OVRO 15\,GHz light-curve. The grey vertical bands mark the different $\gamma$-ray flares.}
         \label{Fig:OVRO_ALL}
   \end{figure}

Figure~\ref{Fig:OVRO_ALL} shows the OVRO 15\,GHz light-curve starting from mid-October 2018 to mid-February 2020. The vertical grey bands marks the time-interval of the \gray{} flares. The light-curve shows the increasing trend of the 15\,GHz flux during the \gray{} observations.
%______________________________________________________________
%______________________________________________________________
%
This Figure clearly shows how the maximum of the 15\,GHz emission occurs about 110\,days after the major \gray{} activity period, followed by a decay ($\sim 4$\,Jy) in 100\,days at a radio flux level comparable with the one at the onset of \gray{} flares. \citet{2010ApJ...722L...7P} found that there is a delay between the \gray{} and the radio emission (the \gray{} emission leads the radio one) up to eight months in the observer's frame (an average of one month in the rest frame). According to~\citet{2010ApJ...722L...7P}, this evidence can be explained by the synchrotron opacity in the nuclear region. Our results, for $z=2.507$, are perfectly in agreement with this scenario.

\begin{figure}
   \centering
   \includegraphics[angle=0,width=\hsize]{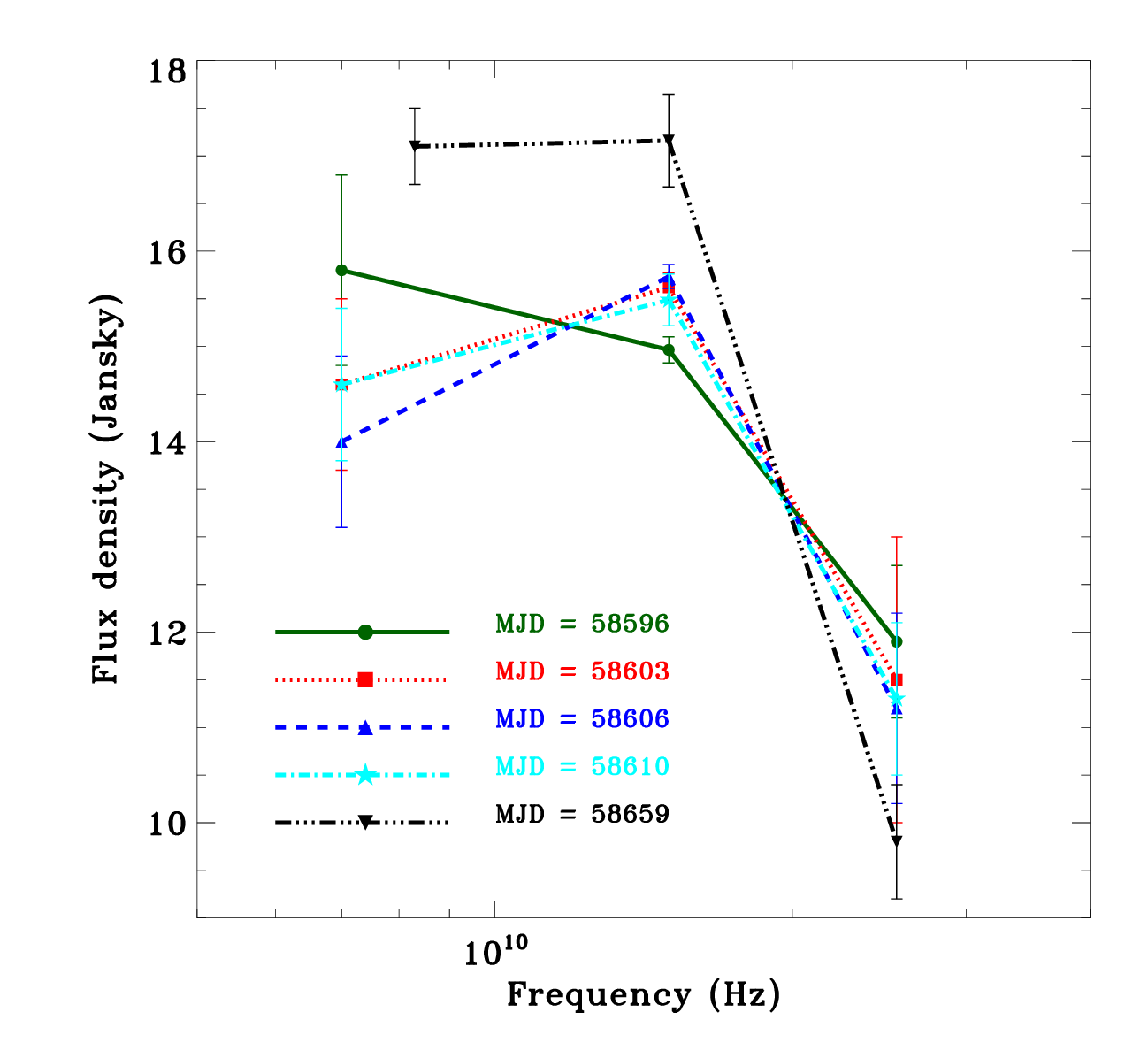}
      \caption{\pks{} radio spectra at different epochs. {We include data from Medicina, OVRO and SRT telescopes at 7, 8.3, 15 and 25.5 GHz. The spectral points were integrated into period ranges whose MJD reference is indicated in the Figure.}}
         \label{FIG:RADIO_BREAK}
   \end{figure}

Figure~\ref{Fig:MWL.LC}, panel (a) shows the SRT/Medicina radio light-curve for the 3 frequency bands observed: K-band (25.5\,GHz) is indicated in red for both telescopes; X-band (8.3\,GHz) in blue and C-band (7\,GHz) in green. Data are reported in Table~\ref{Tab:SRTMED}. The K-band light-curve displays a slight decrease in the observed time range, while lower frequency data seem to show a weak rise, although a clear trend is not evident and a short-term variability is detected. Therefore, a slight spectral break seems to manifest in the 56620-56640 MJD range. As the increasing flux density trend shown by OVRO data at 15\,GHz has extended over about 58700 MJD (see Figure~\ref{Fig:OVRO_ALL}), the observed spectral break is likely to occur in the 15-26\,GHz range.
This break is clearly visible when analysing the SRT and Medicina radio spectra shown in Figure~\ref{FIG:RADIO_BREAK}: a clear break emerges, with a decrease of the flux density at the highest frequency in the observed time range. This is a typical behaviour observed during the course of a flare for extra-galactic jetted sources, which can undergo oscillations and variability of the radio flux density on week/month time scale, in a non-synchronous way at the several radio frequencies~\citep[see, e.g.,][]{2011A&A...531A..95F}.

%
%-------------------------------------------------------------
%-------------------------------------------------------------
%

\begin{figure}
   \centering
   \includegraphics[angle=0,width=9cm]{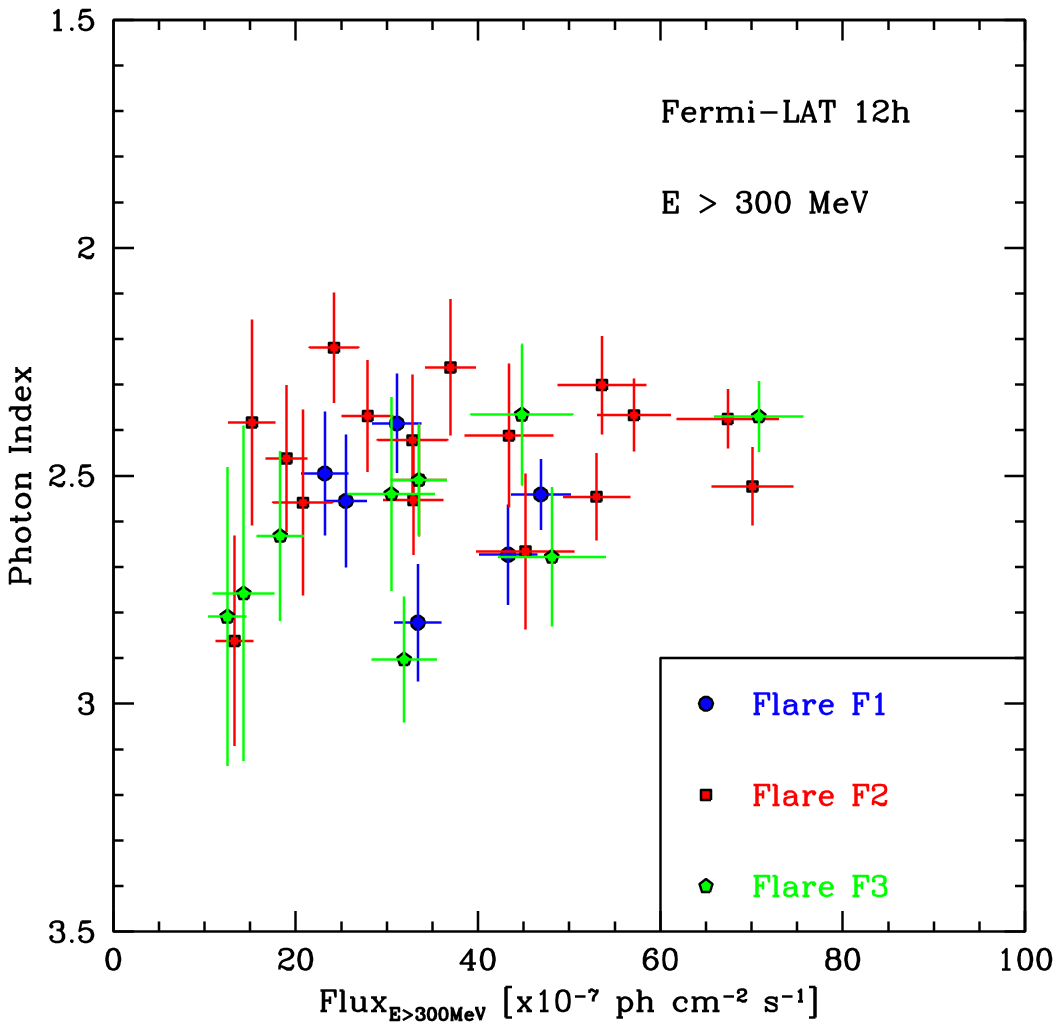}
      \caption{{\it Fermi}-LAT 12h-bin (E>300\,MeV) photon index as a function of the 12h-bin flux during the three main \gray{} flares F1, F2, and F3.}
         \label{Fig:Fermi_FGAMMA_FLARES}
   \end{figure}

%
Figure~\ref{Fig:Fermi_FGAMMA_FLARES} shows the {\it Fermi}-LAT 12h-bin (E>300\,MeV) photon index as a function of the 12h-bin flux during the three main \gray{} flares F1, F2, and F3. It is worth noting that all the flares show the same achromatic behaviour, i.e., the photon index remains almost constant when the flux increases by a factor greater than 5.
The average values of the \gray{} photon indices during the three different flares are $\Gamma_{\rm F1} = 2.58 \pm 0.07$, $\Gamma_{\rm F2} = 2.45 \pm 0.04$, and $\Gamma_{\rm F3} = 2.50 \pm 0.06$. These values are in agreement with the one reported in~\citet{2010ApJ...710.1271A}, $\Gamma = 2.46 \pm 0.18$, for FSRQs, as well as the low scatter as a function of the increasing flux. As noted by~\citet{2010ApJ...710.1271A}, the low dispersion observed may support the idea that a very limited number of physical parameters drive the spectrum shape in the GeV energy range and that it can also be connected to distinct dominant emission mechanisms, e.g. external Compton for FSRQs.
%
%_____________________________________________________________
%                                    One column rotated figure
%-------------------------------------------------------------
%
%_____________________________________________________________
%
%
%

The 12-hour binning of the {\it Fermi}-LAT \gray{} light-curve allows us evaluate the fractional variability and its uncertainty, 
\begin{equation}
F_{\rm var} = \sqrt{\frac{S^{2} - \bar{\sigma^{2}}}{\bar{x}^{2}} } \pm 
\sqrt{\left(\sqrt{\frac{1}{2N}} \frac{\bar{\sigma}^{2}_{\rm err}}{\bar{x}^{2}F_{\rm var}}\right)^{2} +
 \left(\sqrt{\frac{\bar{\sigma}^{2}_{\rm err}}{N}} \frac{1}{\bar{x}}\right)^{2}}
\end{equation}
according to formulae (10) and (B2) in~\cite{2003MNRAS.345.1271V}. We computed $F_{\rm var}$ for both $E>100$\,MeV $E>300$\,MeV in the period MJD 58520--58680, in order to check for any possible dependence of the variability on the energy threshold, obtaining $F_{\rm var}^{\rm E>100MeV} = 0.842 \pm 0.007$ and $F_{\rm var}^{\rm E>300MeV} = 0.723 \pm 0.015$, respectively. These results show a slightly enhanced variability when considering the lower energy threshold with respect to the higher one.
We also computed $F_{\rm var}$ for the data at other frequencies. Only AGILE-GRID ($E>100$\,MeV), {\it Swift}-XRT (0.3-10\,keV), and OVRO (15\,GHz) data yield non-null $F_{\rm var}$ values: $F_{\rm var}^{\rm AGILE} = 0.62 \pm 0.04 $,  $F_{\rm var}^{\rm XRT} = 0.09 \pm 0.05 $, and $F_{\rm var}^{\rm OVRO} = 0.078  \pm  0.003$, respectively. We note that, as seen in other blazars~\citep[e.g., 3C 454.3,][]{2010ApJ...712..405V}, $F_{\rm var}$ is higher in the \gray{} energy band than in the radio band. We should also consider that the calculation of $F_{\rm var}$ could be influenced both by the binning of the light-curve and by the source coverage at different frequencies, as discussed in detail in \cite{2019Galax...7...62S}.
%

%
%_____________________________________________________________________
% DOUBLING/HALVING times
%----------------------------------------------------------------------
\begin{table*}
\caption{Minimum doubling (R, rising portion of the light-curve) and halving (D, decaying portion of the light-curve) time $\tau_{d}$. }     
\label{TAB:doubling}      
\centering    
\begin{tabular}{ccccccc}        
\hline        
$t_{1}$ &  $t_{2}$  &  $F(t_{1})$  &  $F(t_{2})$  & $\tau_{\rm d}$  & Doubling/Halving & Significance \\
(MJD)   &  (MJD)   &  ($10^{-7}$\phcmsec)  & ($10^{-7}$\phcmsec) & (Days) & Raise/Decay &  ($\sigma(\tau_{d})$) \\
\hline
58539.75 &  58540.25  &   $6.9\pm1.9$   &   $36.0\pm4.6$  &  0.21  & R  &  15.3 \\
58527.75 &  58528.25  &   $34.0\pm4.6$  &  $8.2\pm1.9$    &  0.25  & D  &  5.6 \\ 
\hline
\end{tabular}
\end{table*}
We can also estimate the minimum variability time scale for $E>100$\,MeV by analysing the 12\,hr-binned {\it Fermi}-LAT light-curve,  $t_{\rm var} = {\rm ln}(\rm2) \times \tau_{\rm d}$\,days, where  $\tau_{\rm d}$ is the doubling(R, rising portion of the light-curve)/halving(D, decaying portion of the light-curve) time defined by 
\begin{equation}\label{EQ:tau_d}
F(t_{2}) = F(t_{1}) \times 2^{(t_{2} - t_{1})/\tau_{\rm d}},
\end{equation}
and $F(t_{1})$ and $F(t_{2})$ are the $E>100$\,MeV \gray{} fluxes at the times $t_{1}$ and $t_{2}$, respectively. The AGILE-GRID \gray{} data have a much larger binning (48\,hr) compared with the {\it Fermi}-LAT one (12\,hr), which makes the AGILE data less constraining to assess time variability on short timescales.
Table~\ref{TAB:doubling} shows the minimum doubling/halving times and their significance, $\sigma(\tau_{\rm d}) = |F(t_{1}) - F(t_{2})|/\sigma(F(t_{1}))$. We selected those doubling/halving times with $\sigma(\tau_{d}) \ge 3$. Assuming $\tau_{\rm d} = \min\{\tau_{\rm d}(R);\tau_{\rm d}(D) \}$, we obtain the minimum variability timescale $t_{\rm var} = 0.15$\,days. 

This quantity can be used to derive the minimum Doppler factor~\citep{1995MNRAS.273..583D}, 
\begin{equation}\label{EQ:delta_min}
\delta_{\rm min} \ge \left[  3.5\times10^{3} \frac{(1+z)^{2\alpha}(1+z-\sqrt{1+z})^{2}F_{x}(3.8\nu_{\gamma}\nu_{x})^{\alpha}}{t_{\rm var}}  \right]^{1/(4+2\alpha)},
\end{equation}
where $z$ is the source redshift, $\alpha$ is the energy spectral index in the X-ray band, $F_{x}$ is the X-ray flux at 1\,keV in $\mu$Jy, $\nu_{\gamma}$ is the average energy of the maximum energy bin in GeV of the \gray{} spectrum, and $\nu_{x}$ is 1\,keV. In order to derive the X-ray flux and spectral properties, we stacked all the {\it Swift}/XRT observations, because of their moderate variability. We obtain $F_{\rm 0.3-10\,keV} = (1.65^{+0.04}_{-0.04})\times10^{-11}$\,\ferg  and $\alpha = 1.32^{+0.06}_{-0.05}$, which yields $\delta_{\rm min} \ge 24.3$. 

%%%%%%%%%%%%%%%%%%%%%%%%%%%%%%%%
\subsection{Spectral energy distribution}\label{Disc:SED}
\begin{figure}
   \centering
   \includegraphics[angle=0,width=9cm]{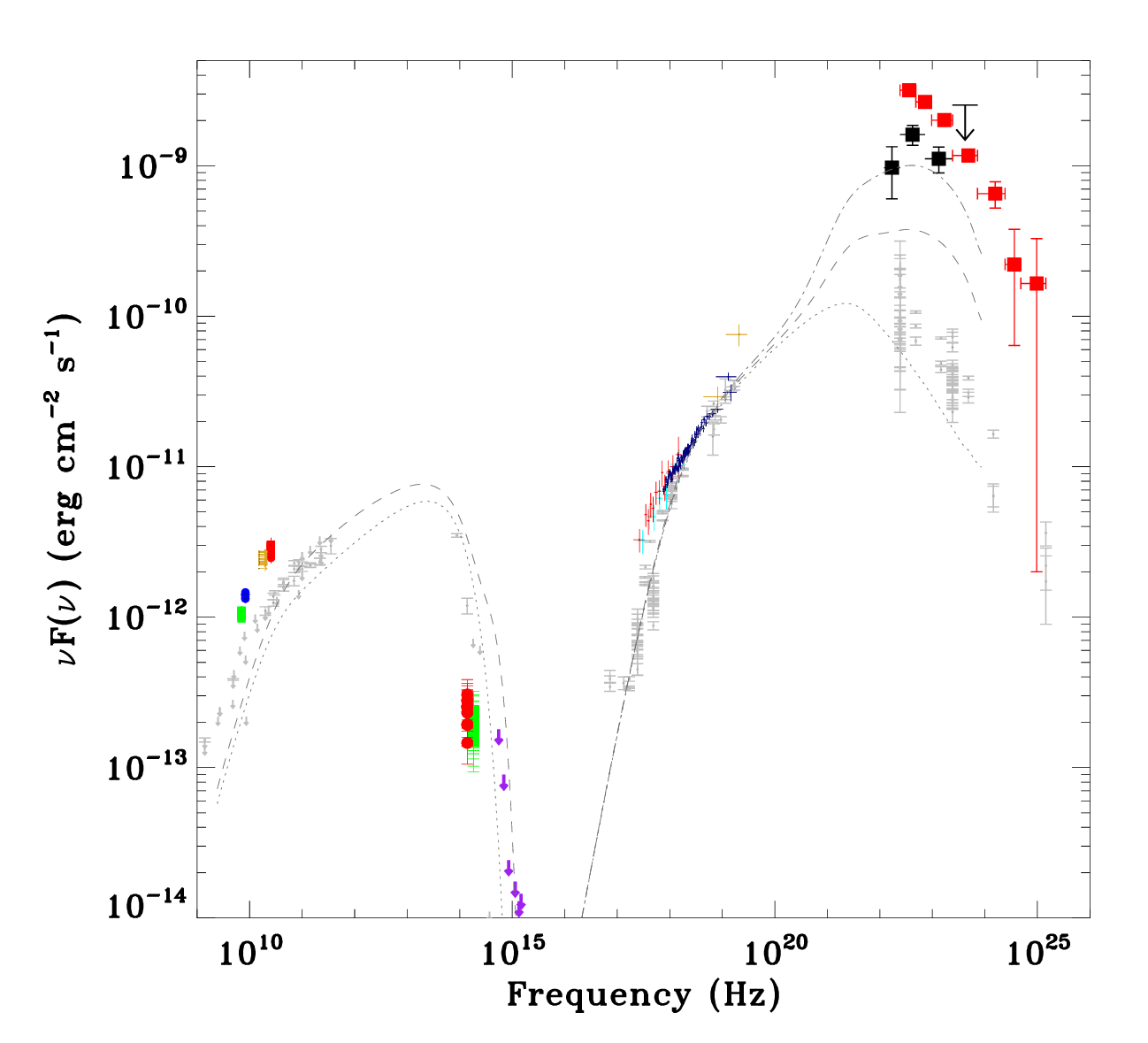}
     \caption{\pks{} spectral energy distributions for the three major \gray{} flares (see Section~\ref{Disc:SED} for details). 
     In the radio band, green, blue, golden and red points  represent 7\,GHz, 8.3\,GHz, 15\,GHz, and 25.5\,GHz data. In the IR band, green and red points represent REM-H and REM-K bands, respectively. Purple upper limits represent the {\it Swift}/UVOT data, integrated during the whole observing campaign. In the X-ray energy band, cyan and red points represent {\it Swift}/XRT spectra accumulated on MJD~58568--58578 (\gray{} F1) and MJD~58590--58606 (\gray{} F2$+$F3), respectively. Blue points represent {\it NuSTAR} data acquired on MJD~58550, while golden points are the sum of all the INTEGRAL/IBIS observations. 
For sake of clarity, in the \gray{} energy band we report data pertaining to flare F1. Red filled squares correspond to {\it Fermi}-LAT F1 data, while black filled squares correspond to AGILE-GRID data acquired on a similar period.
     Small grey points are archival data.
     The dotted, dashed and dot-dashed dark-grey lines correspond to the 2010 SEDs for the average $\gamma$-ray state, the 1-month integration, and the 5-day \gray{} flare, respectively, as described in~\citet{2011ApJ...736L..30D}.
     }
         \label{Fig:PKS1830sed}
   \end{figure}

%%%%%%%%%%%%%%%%%%%%%%%%%%%%%%%%
%
During our observing campaign we collected multi-wavelength data covering the main \gray{} flares. Figure~\ref{Fig:PKS1830sed} shows the \pks{} spectral energy distribution (SED). In the radio and IR energy bands, points and colours follow those presented in Figure~\ref{Fig:MWL.LC}. Purple upper limits represent the {\it Swift}/UVOT data, integrated during the whole observing campaign. In the X-ray energy band, cyan and red points represent {\it Swift}/XRT spectra accumulated on MJD~58568--58578 (\gray{} F1) and MJD~58590--58606 (\gray{} F2$+$F3), respectively. Blue points represent {\it NuSTAR} data acquired on MJD~58550, while golden points are the sum of all the INTEGRAL/IBIS observations. The combined X-ray spectral model fitting was performed with the following parameters: {\tt const*phabs*(zphabs*pow)} fixing the absorption along the line-of-sight to $N_{\rm H}^{\rm gal} = 0.187 \times 10^{22}$\,cm$^{-2}$ and $N_{\rm H}^{\rm lens} = 3.1 \times 10^{22}$\,cm$^{-2}$ (assuming $z_{\rm lens}=0.89$). Both fits yield a photon index of $1.46\pm0.01$. 
%
%
%_____________________________________________________________
%                                    One column rotated figure
%-------------------------------------------------------------
%
%_____________________________________________________________
%

In the \gray{} energy band we investigated the {\it Fermi}-LAT finer time-binning in order to select the most significant flare episodes, F1, F2 and F3. 
The corresponding AGILE-GRID photon indices are $\Gamma(\rm F1) = 2.35 \pm 0.14$,  $\Gamma(\rm F2) = 2.29 \pm 0.09$, and  $\Gamma(\rm F3) = 2.00 \pm 0.16$, respectively. Small grey points are archival data provided by the ASI/SSDC {\it SED Builder Tool}~\citep{2011arXiv1103.0749S} which include public catalogs and surveys. The data show the typical double-humped shape of the blazar SED. Moreover, while the rising branch of the inverse Compton and the poorly constrained synchrotron emission are almost consistent with the previous SEDs, the high-energy peak ($E>100$\,MeV)  is about a factor of 3--4 more intense with respect to the flare discussed in~\cite{2011ApJ...736L..30D}, whose SED fits are reported as dotted line (quiescent state), dashed line (one-month integration around the 2010 flare) and dot-dashed line (5-day flare).

We model the flare F1 for which we have simultaneous data in radio, IR, optical, X-ray and $\gamma$-ray bands (IR data are relevant to constrain the synchrotron component). The data show very high Compton dominance, with stronger daily variability in $\gamma$-rays than the others bands: these data challenge a simple one-zone model~\cite[see e.g.,][]{2017ApJ...843L..23V}. 
Adapting the original model discussed in~\cite{2009ApJ...706.1433V}, we first consider the emission in optical-UV. Assuming a magnification factor due to gravitational lensing of the order of 10~\citep[see e.g.][]{2011ApJ...736L..30D}, the accretion disk has to radiate $L_{\rm d} \lesssim 10^{45}$\,\ergsec at black-body temperature $T_{\rm d}\approx
3\times 10^{4}$K, while the broad-line region (BLR) reprocesses 5$\%$ of this radiation from a radius $R_{\rm BLR}\approx 0.05$\,pc, typical for these disc
luminosities. We also consider a dusty torus having extension $R_{\rm Torus}\approx 1$\,pc that emits infrared photons at black-body
temperature $T_{\rm Torus}\approx 100$\,K with luminosity $L_{\rm T}\simeq L_{\rm d}$.
Therefore, we consider an internal electron population {\it cI} in a jet region of longitudinal size $L\simeq 10^{17}$\,cm and tangled magnetic field $B\simeq 1$\,Gauss, moving with bulk Lorentz factor $\Gamma \simeq 18$ \citep[see, e.g.,][]{1998ApJ...509..608T}, in which the emission is due to synchrotron process and inverse Compton with the same synchrotron photons plus external photons coming from the accretion disk, the BLR and the dusty torus. At the BLR edge we assume that the cooled electrons of {\it cI} are re-accelerated to a {\it cII} component (of size $R\simeq 3\times10^{16}$\,cm) by, e.g., kink or tearing instability~\citep{2022MNRAS.510.2391B} that slightly modifies the viewing angle $\theta$. This second component accounts for the enhanced $\gamma$-ray flux via inverse Compton with the external soft photons. The magnetic field in {\it cII} is then assumed to decrease to 0.2\,Gauss as the plasmoid moves away from the center, towards the BLR edge. Moreover, we assume the emitters to have a jet-frame distribution of the random energies ($\gamma m c^2$), starting from $\gamma_{\rm min} = 30$~\citep[see, e.g.,][]{2011ApJ...736L..30D}, in the form of a standard broken power-law
\begin{equation}
\label{eq:ne_gamma}
n_e(\gamma)=\frac{K\,\gamma_b^{-1}}{
(\gamma/\gamma_b)^{\zeta_{1}}+(\gamma/\gamma_b)^{\zeta_{2}}},
\end{equation}
where $\zeta_{1}$ and $\zeta_{2}$ are the spectral indices for $\gamma<\gamma_b$ and $\gamma>\gamma_b$, respectively, $\gamma_b$ is the Lorentz factor at the break and the normalisation is assumed $K\simeq 1$ corresponding to an electron density $\sim 70$\,cm$^{-3}$ for {\it cI} and $K\simeq 30$ corresponding to an electron density $\sim 10^{3}$\,cm$^{-3}$ for {\it cII}, respectively.
Table~\ref{pks1830:tab:AGN} shows the parameters we assumed for \pks{}, while
%
%_____________________________________________________________
%                                    TABLE FOR THE AGN PARAMETERS
%-------------------------------------------------------------
%
\begin{table}
\begin{center}
\caption{Model parameters for the AGN.} \label{pks1830:tab:AGN}
\begin{tabular}{ lccc }
\hline \noalign{\smallskip}
  {Ext. source} & ${L(10^{45}\,\ergsec)}$ & {$T$(K)} & ${R}$(pc) \\
  \hline
  {\it Accr. disk}  & 1 & $3\times 10{^4}$   & $R_{\rm BLR}=0.05$     \\
  {\it Dusty torus} & 1 & 100            & $R_{\rm Torus}= 1$     \\
\noalign{\smallskip} \hline
\end{tabular}
\end{center}
\end{table}
Table~\ref{pks1830:tab:sedfit} shows the values of the parameters for the two components, {\it cI} and {\it cII}, responsible for the overall SED. Assuming a bulk Lorentz factor $\Gamma \simeq 18$ and the values of the viewing angles $\theta$ reported in Table~\ref{pks1830:tab:sedfit}, we obtain Doppler factors of the order of 29--32 for  {\it cI} and {\it cII}, respectively. These values are in agreement with the minimum Doppler factor derived in Section~\ref{Disc:MWL}.
%
%
%_____________________________________________________________
%                                    TABLE FOR THE MODEL PARAMETERS
%-------------------------------------------------------------
\begin{table}
\begin{center}
\caption{Model parameters for the \gray{} flare F1.} \label{pks1830:tab:sedfit}
\begin{tabular}{ lcccccc }
\hline
\noalign{\smallskip}
  {Comp.} & ${\theta}$ & {$B$(G)} & ${\gamma_b}$ & ${\zeta_1}$ & ${\zeta_2}$ & ${\gamma_{\rm max}}$ \\
  \hline
  {\it cI} & 1.5 & 1   &  600 & 2.3 & 3.5 & $10^{3}$ \\
  {\it cII} & 1 & 0.2  &  500 & 2.1 & 3.5 & $3\times10{^3}$ \\
\noalign{\smallskip} \hline
\end{tabular}
\end{center}
\end{table}

%
%_____________________________________________________________
%                                    One column rotated figure
%-------------------------------------------------------------
   \begin{figure}
   \centering
   \includegraphics[angle=0,width=9cm]{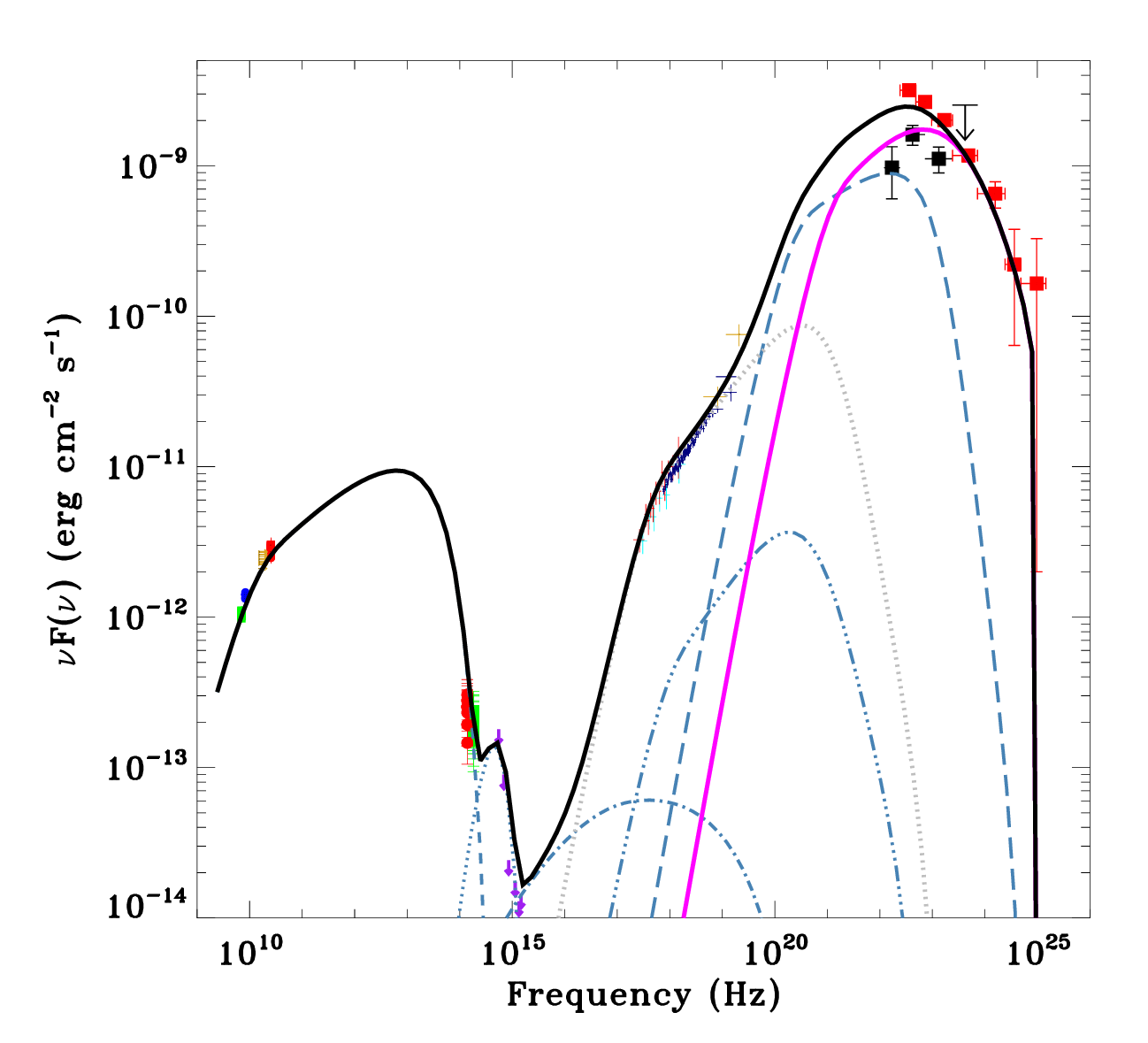}
      \caption{\pks{} spectral energy distributions for the flare F1 with different fit components. Data points follow the color scheme adopted for Figure~\ref{Fig:PKS1830sed}. Different lines correspond to different spectral components. Blue-dotted is the disk emission, blue-dashed is the synchrotron emission, blue-dashed-dotted is the synchrotron self-Compton emission, blue-triple-dotted-dashed is external Compton off the disk photons emission, blue-long-dashed is the external Compton off the BLR photons emission, grey-dotted is the external Compton off the torus photons emission, magenta-solid is the external Compton off the BLR photons emission of component $cII$, and black-solid is the sum of all the different components. }
         \label{Fig:PKS1830sedFIT}
   \end{figure}
%
%_____________________________________________________________
%
%
Figure~\ref{Fig:PKS1830sedFIT} shows the multi-component fit to our data, as described above.  The different lines represent different emission components. We note that the main contribution to the IC peak is provided by the inverse Compton off the BLR photons, as suggested by the achromatic behaviour reported in Figure~\ref{Fig:Fermi_FGAMMA_FLARES}.

The Eddington luminosity is $L_{\rm Edd} \approx 6.2\times 10^{46}$\,\ergsec{}, where we assumed the value of the black hole mass reported by~\cite{2005MNRAS.362.1157N}, $M_{\rm BH} = 5\times 10^{8}$\,M$_{\odot}$. 
The total power carried in the jet, $P_{\rm jet}$, can be calculated following \citet{2001MNRAS.327..739G} as
\begin{equation}
P_{\rm jet} = P_{\rm B} + P_{\rm p} + P_{\rm e} + P_{\rm rad},
\label{eq:Pjet}
\end{equation}
where $P_{\rm B}$, $P_{\rm p}$, $P_{\rm e}$, and $P_{\rm rad}^{\rm bol}$ are the power carried by the magnetic field, the cold protons, the relativistic electrons, and the produced radiation, respectively. In order to compute the different components, we use the formalism presented in \citet{2008MNRAS.385..283C}. We obtain:
$P_{\rm B} \approx 8\times10^{45}$\,erg\,s$^{-1}$, 
$P_{\rm e} \approx 5\times10^{44}$\,erg\,s$^{-1}$,
$P_{\rm p} \approx 2\times10^{46}$\,erg\,s$^{-1}$, 
$P_{\rm rad} \approx 2.0\times10^{46}$\,erg\,s$^{-1}$, 
which yields $P_{\rm jet} \approx 5\times10^{46}$\,erg\,s$^{-1}$.
This is comparable to the maximum $P_{\rm jet}$ value computed during the October 2010 flare in~\cite{2011ApJ...736L..30D}. We also note that, due to the high variability of \pks{}, the comparison of model parameters are not always straightforward. While the \gray{} flare modelled in~\cite{2011ApJ...736L..30D} is not dramatically different from the one discussed in this work, the SED discussed in~\cite{2005A&A...438..121D} reached a \gray{} peak more than two order of magnitude lower than the present one, since they reported the Third EGRET Catalogue~\citep{1999ApJS..123...79H} spectrum. We also note that, despite a different definition of the different \gray{} flares and SED model parameters, our $P_{\rm jet}$ estimate is compatible with the one reported in~\cite{2021ApJ...915...26A}.
%

%%%%%%%%%%%%%%%%%%%%%%%%%%%%%%%%
\subsection{Prospects for detection in the MeV and VHE bands}\label{Disc:TEV}
\begin{figure}
   \centering
   \includegraphics[angle=0,width=9cm]{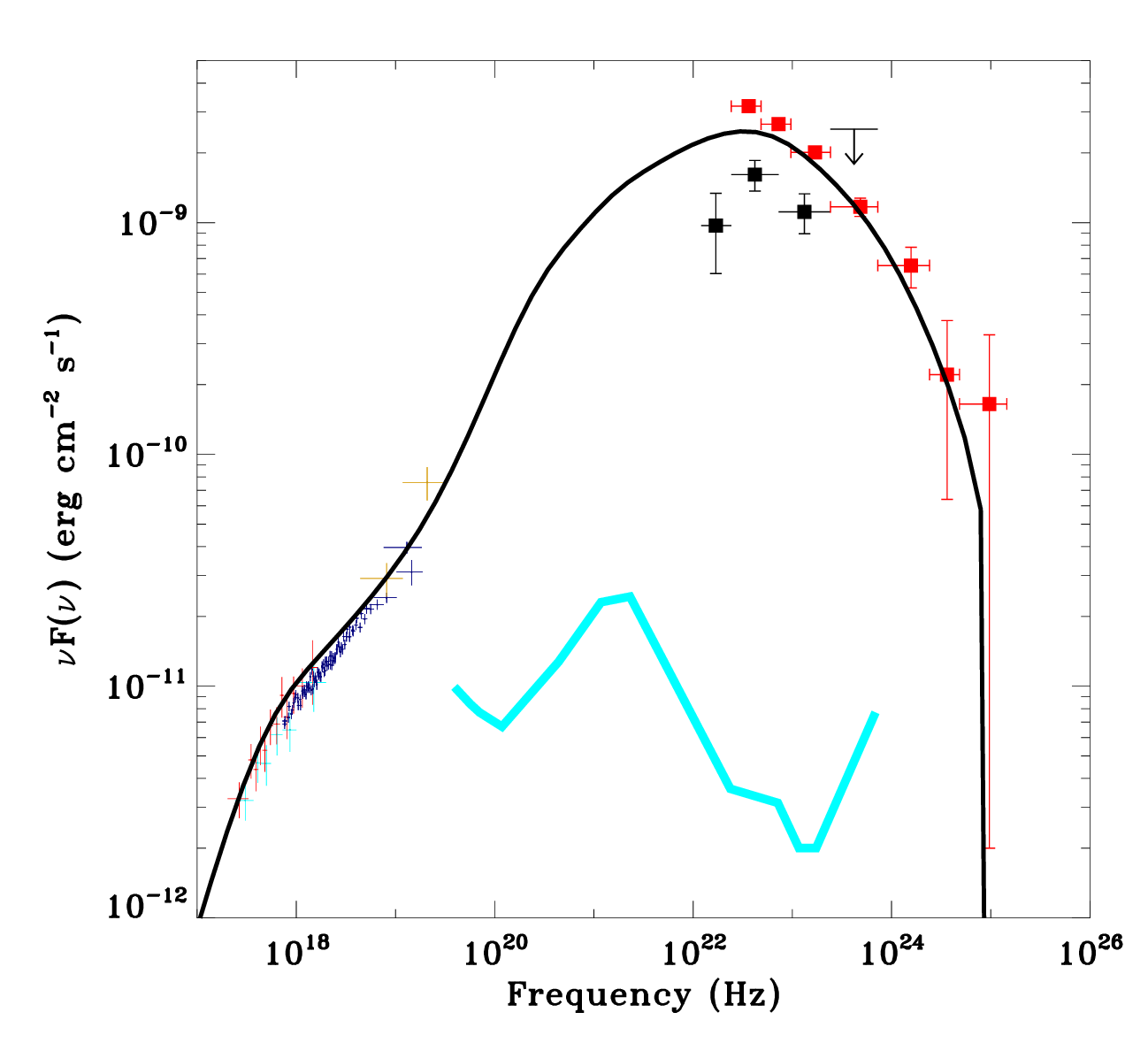}
      \caption{Inverse Compton peak region of \pks{}. Data points and the black-solid line follow the color scheme adopted for Figure~\ref{Fig:PKS1830sed}.  The cyan curve represents the ASTROGAM sensitivity for an integration time of 6 days.}
         \label{Fig:MeV}
   \end{figure}

%%%%%%%%%%%%%%%%%%%%%%%%%%%%%%%%
%
Figure~\ref{Fig:MeV} shows the inverse Compton peak region of \pks{} data. The cyan curve represents the ASTROGAM sensitivity for an integration time of 6 days~\citep[see][for further details]{2019A&A...621A..82V}. ASTROGAM is a proposed Observatory space mission dedicated to the study of the non-thermal Universe in the photon energy range from 0.3 MeV to 3 GeV \citep{2017ExA....44...25D, 2021ExA....51.1225D}. ASTROGAM could be a very useful mission in order to study high-redshift, powerful FSRQs such as \pks{}~\citep[][and this work]{2011ApJ...736L..30D}, 4C~$+$71.07~\citep{2019A&A...621A..82V} and a small sample in~\citet{2020ApJ...889..164M}. As seen in Figure~\ref{Fig:MeV}, ASTROGAM would sample the IC component, providing crucial data which cover the energy range from 0.1 MeV up to a few hundreds of MeV. This would allow us to cover a currently unsampled energy range, and to obtain important data to constrain the IC peak in high-z FSRQs. 
%_____________________________________________________________
%                                    One column rotated figure
%-------------------------------------------------------------
%
%_____________________________________________________________
%

\begin{figure}
   \centering
   \includegraphics[angle=90,width=9cm]{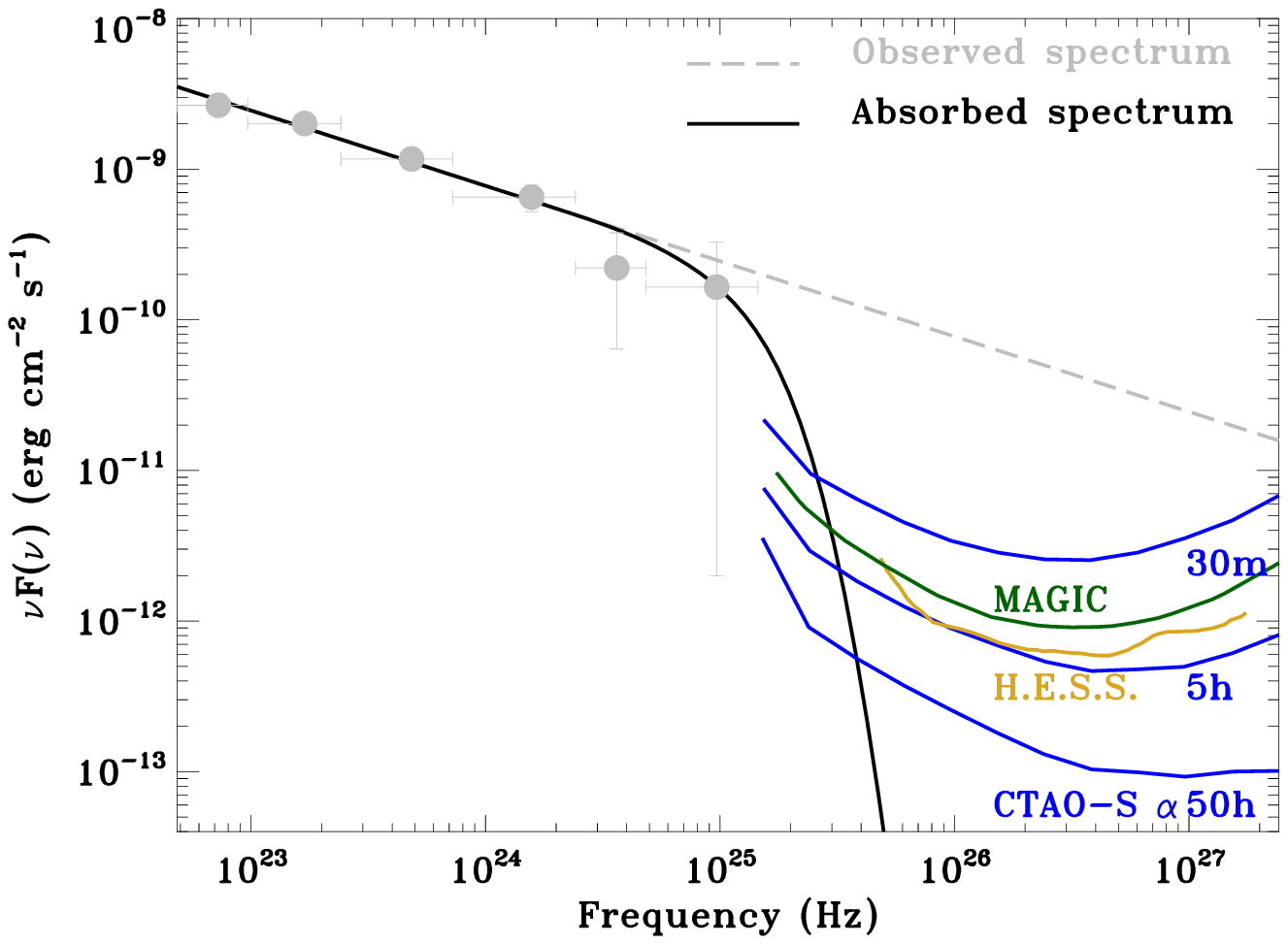}
      \caption{Grey data and dashed line represent the {\it Fermi}-LAT data and power-law spectrum during flare F1, while the black solid line represent the EBL-absorbed \pks{} spectrum at energies above 10\,GeV. Green and golden lines represent the MAGIC and H.E.S.S.-I differential sensitivity curves for and integration time of 50\,hr, respectively. Blue lines represent the CTAO-S Alpha Configuration differential sensitivity curves for different integration times (0.5, 5 and 50)\,hr (See \href{https://www.cta-observatory.org/science/ctao-performance/}{https://www.cta-observatory.org/science/ctao-performance/}).}
         \label{Fig:VHE}
   \end{figure}

Figure~\ref{Fig:VHE} shows the {\it Fermi}-LAT data and power-law spectrum during flare F1, extrapolated to higher energies, in the optimistic assumption that no intrinsic cut-off applies. The correction for absorption by the extra-galactic background light (EBL), providing substantial attenuation only above a few tens of GeV, has been applied using the model of~\citet[][black solid line]{2011MNRAS.410.2556D}.
%_____________________________________________________________
%                                    One column rotated figure
%-------------------------------------------------------------
%
%_____________________________________________________________
%
We note that during a previous \pks{} \gray{} flare that occurred on 2014 July 27 \citep{2014ATel.6361....1K}, the source was also observed by the H.E.S.S.-II array, starting about 20 days after the {\it Fermi}-LAT alert in order to investigate possible delayed emission at very high energies~\citep{2019MNRAS.486.3886H}. This observation was performed by adding the large CT5 telescope, which provides detection capabilities below 100\,GeV~\citep{2017A&A...600A..89H}. No significant signal was detected above $\sim 67$\,GeV. 
The green, gold, and blue solid lines correspond to the MAGIC (50\,hr), H.E.S.S (50\,hr), and CTAO alpha configuration\footnote{The ``Alpha Configuration'' for the southern CTAO array consists of 14 medium-sized telescopes and 37 small-sized telescopes. See \href{https://www.cta-observatory.org/science/ctao-performance/}{https://www.cta-observatory.org/science/ctao-performance/} for further details.} (50, 5, and 0.5\,hr) differential sensitivity, respectively. MAGIC can observe \pks{} at a Zenith angle of about 50\,deg, which increases its low energy threshold, while the H.E.S.S.-I sensitivity (when considering the usual CT1--4 configuration) does not extend significantly at energies lower than 100\,GeV. CTAO, in its alpha configuration, might be able to detect \pks{} with a short integration time (0.5--5)\,hr up to 100\,GeV. 
The detection of FSRQs by IACTs may challenge the current paradigm of the HE-VHE emission occurring within or at the edge of the BLR. As shown in~\citet{2018MNRAS.477.4749C}, the \gray{} emission from {\it Fermi}-LAT FSRQs might be explained by means of emitting mechanisms which do not involve the jet interaction with the BLR. Moreover, evidence is emerging that for blazars the location of the \gray{} emitting region may not always be placed at the same distance from the central black-hole during different flaring episodes of the same source as suggested by~\citet{2013MNRAS.431..824B} for PKS~1510$-$089 or by~\citet{2016MNRAS.458..354C} and~\citet{2016ApJ...830...94F} for 3C 454.3. A few FSRQs have been already detected by current IACTs (see also \href{http://tevcat.uchicago.edu/}{http://tevcat.uchicago.edu/}). Therefore, the detection of \pks{} by CTAO during particular strong flares could shed light on the location of the gamma-ray emitting zone and the related emission mechanisms.

%%%%%%%%%%%%%%%%%%%%%%%%%%%%%%%%
\section{Summary and Conclusions}
%%%%%%%%%%%%%%%%%%%%%%%%%%%%%%%%
%
In this paper we presented the multi-wavelength behaviour, from radio frequencies up to the $\gamma$-ray energy band, of the lensed quasar \pks{} during multiple flaring episodes that occurred in the period mid-February to mid-April 2019 through nearly simultaneous observations presented here for the first time. We can summarise our findings as follows:

\begin{enumerate}
 \item The {\it Fermi}-LAT data show three major \gray{} flares, F1 (MJD 58575.2--58576.1), F2 (MJD 58595.0--58598.8), and F3 (MJD 58601.5--58603.4), respectively. The minimum variability timescale for $E>100$\,MeV is $t_{\rm var} = 0.15$\,days, while it emerges  there could be a slightly enhanced fractional variability when considering the lower energy threshold ($E>100$\,MeV) with respect to the higher one ($E>300$\,MeV). Moreover, we confirm the higher fractional variability in the \gray{} energy band with respect to the one at lower frequencies. Another interesting result is on \gray{} spectral variability as a function of the flux. As already noted for other sources \citep[3C~454.3,][in the 0.2-10\,keV energy band]{2014styd.confE.167V} there is a roughly achromatic increase of the \gray{} emission. This could be explained, for example, with the dominance of the external Compton emission mechanism in the \gray{} energy band.
 \item X-ray data show moderate variability during the whole observing period. A detailed analysis of {\it Chandra} data show a hint of variability of the absorbing column density in the lensing galaxy.
 \item Radio data show an interesting behaviour. Investigating the data in the frequency range 7--25.5\,GHz we find a spectral break above 15\,GHz, with a decrease of the flux density at the highest frequency in the observed time range. Similar breaks in the radio spectra could be found in other extra-galactic jetted sources, due to radio flux variability at different wavelengths in times. Moreover, 15\,GHz data show a continuous flux rising up to a maximum occurring about 110\,days after the \gray{} flares, suggesting possible different locations of the \gray{} and radio emission zones.
 \item UV-optical data are challenging to be acquired. While we have only upper-limits in the UV-optical wavebands, infra-red data show no particular variability pre-, during, and post-flare F1 episode.
\item The SED modelling shows that our data are consistent with a multiple-component emission model, where the emission in the energy band above 100\,MeV could be interpreted by the inverse Compton emission, at the edge of the BLR, of electrons re-accelerated by kink or tearing instability. Moreover, the total jet power is comparable to that of the 2010 flare.
\item Finally, \pks{} is an excellent candidate for upcoming both Compton and VHE facilities. Upcoming Compton missions will probe the IC peak in high-z FSRQs while a possible detection of \pks{} would increase the number of FSRQs detected at VHE and provide useful information of the location of the \gray{} emitting zone during different flares.
\end{enumerate}

%%%%%%%%%%%%%%%%%%%%%%%%%%%%%%%%
\section*{Acknowledgements}
%%%%%%%%%%%%%%%%%%%%%%%%%%%%%%%%
%
S.V., I.D., C.P., F.C., A.dR., L.dG., S.K., M.N.I., A.P.P., E.E., L.P., G.P., S.P., S.R., G.V., F.V., V.V. contributed equally to this work. We thank the referee for the prompt reply and the valuable comments which improved the quality of the manuscript. The authors acknowledge financial contribution from the grant ASI I/028/12/0. SV acknowledges financial contribution from the agreement ASI--INAF n.2017-14-H.0.
The Sardinia Radio Telescope is funded by the Ministry of University and Research (MIUR), Italian Space Agency (ASI), and the Autonomous Region of Sardinia (RAS) and is operated as National Facility by the National Institute for Astrophysics (INAF). The Medicina radio telescope is funded by the Ministry of University and Research (MIUR) and is operated as National Facility by the National Institute for Astrophysics (INAF).
 This research has made use of data from the OVRO 40-m monitoring program~\citep{2011ApJS..194...29R}, supported by private funding from the California Insitute of Technology and the Max Planck Institute for Radio Astronomy, and by NASA grants NNX08AW31G, NNX11A043G, and NNX14AQ89G and NSF grants AST-0808050 and AST-1109911.
S.K. acknowledges support from the European Research Council (ERC) under the European Unions Horizon 2020 research and innovation programme under grant agreement No.~771282.
Part of this work is based on archival data, software or online services provided by the Space Science Data Center -- ASI.

%%%%%%%%%%%%%%%%%%%%%%%%%%%%%%%%
\section*{Data Availability}
%%%%%%%%%%%%%%%%%%%%%%%%%%%%%%%%
%
The data underlying this article are publicly available from the {\it Fermi}-LAT, AGILE-GRID, INTEGRAL, {\it NuSTAR}, {\it Chandra}, {\it Swift}, SRT/Medicina, and REM archives and processed with publicly available software (SRT/Medicina SDI software can be available upon request). OVRO 40-m data are available upon request to Sebastian Kiehlmann (skiehl@physics.uoc.gr).

%%%%%%%%%%%%%%%%%%%% REFERENCES %%%%%%%%%%%%%%%%%
\bibliographystyle{mnras}
%%%%%%%%%\bibliography{PKS1830-211} % if your bibtex file is called example.bib

%%%%%%%%%%%%%%%%%%%%%%%%%%%%%%%%%%%%%%%%%%%%%%%%%%

%%%%%%%%%%%%%%%%%%%%%%%%%%%%%%%%%%%%%%%%%%%%%%%%%%
%                                                       APPENDICES                                                                            %
%%%%%%%%%%%%%%%%%%%%%%%%%%%%%%%%%%%%%%%%%%%%%%%%%%

%%%%%%%%%%%%%%%%%%%%%%%%%%%%%%%%%%%%%%%%%%%%%%%%%%
%%%%%%%%%%%%%%%%%%%%%%%%%%%%%%%%%%%%%%%%%%%%%%%%%%
%%%%%%%%%%%%%%%%%%%%%%%%%%%%%%%%%%%%%%%%%%%%%%%%%%
%%%%%%%%%%%%%%%%%%%%%%%%%%%%%%%%%%%%%%%%%%%%%%%%%%
%%%%%%%%%%%%%%%%%%%%%%%%%%%%%%%%%%%%%%%%%%%%%%%%%%
%%%%%%%%%%%%%%%%%%%%%%%%%%%%%%%%%%%%%%%%%%%%%%%%%%
%%%%%%%%%%%%%%%%%%%%%%%%%%%%%%%%%%%%%%%%%%%%%%%%%%
%%%%%%%%%%%%%%%%%%%%%%%%%%%%%%%%%%%%%%%%%%%%%%%%%%

\appendix

%%%%%%%%%%%%%%%%%%%%%%%%%%%%%%%%
\section{Gamma-ray Observations}\label{SEC:obs:gamma}
%%%%%%%%%%%%%%%%%%%%%%%%%%%%%%%%

%%%%%%%%%%%%%%%%%%%%%%%%%%%%%%%%
\subsection{AGILE data}\label{AGILEdata}\label{SEC:obs:gamma:agile}
%%%%%%%%%%%%%%%%%%%%%%%%%%%%%%%%
%
%
The AGILE satellite \citep{2009A&A...502..995T} is a mission of the Italian Space Agency (ASI) devoted to high-energy astrophysics. 
The AGILE scientific instrument combines four active detectors yielding broad-band coverage from hard X-ray to \gray{} energies: a Silicon Tracker~\citep[ST;][30~MeV--50~GeV]{2003NIMPA.501..280P}, a co-aligned coded-mask hard X-ray imager, Super--AGILE \citep[SA;][18--60~keV]{2007NIMPA.581..728F}, a non-imaging CsI Mini--Calorimeter~\citep[MCAL;][0.3--100~MeV]{2009NIMPA.598..470L}, and a segmented Anti-Coincidence System~\citep[ACS;][]{2006NIMPA.556..228P}. Any \gray{} detection is obtained by the combination of ST, MCAL and ACS; these three detectors form the AGILE Gamma-Ray Imaging Detector (GRID). A ground segment alert system allows the AGILE team to perform the full AGILE-GRID data reduction and the preliminary quick-look scientific analysis \citep{2013NuPhS.239..104P, 2014ApJ...781...19B,2019RLSFN..30S.217P}.

\pks{} underwent an exceptionally bright active phase in $\gamma$-rays which started at the end of February 2019 and lasted approximately 2 months, as preliminarily reported in~\cite{2019ATel12541....1L, 2019ATel12594....1P, 2019ATel12603....1P, 2019ATel12601....1A, 2019ATel12622....1C}.
We carried out the analysis of the {\grid{} consolidated data (archive \verb+ASDCSTDk+) above 100 MeV with the new \verb+Build_25+ scientific software, \verb+FM3.119+ calibrated filter, \verb+H0025+ response matrices.} We applied South Atlantic Anomaly event cuts and $80^{\circ}$ Earth albedo filtering. Only incoming \gray{} events with an off-axis angle lower than $60^{\circ}$ were selected for the analysis. Statistical significance and flux determination of the point sources were calculated using the AGILE multi-source likelihood analysis software \citep[MSLA;][]{2012A&A...540A..79B} based on the Test Statistic (TS) method as formulated by \cite{1996ApJ...461..396M}. This statistical approach provides a detection significance assessment of a \gray{} source by comparing maximum-likelihood values of the null hypothesis (no source in the model) with the alternative hypothesis (point source in the field model).
 
\begin{figure}
   \centering
   \includegraphics[angle=0,width=9cm]{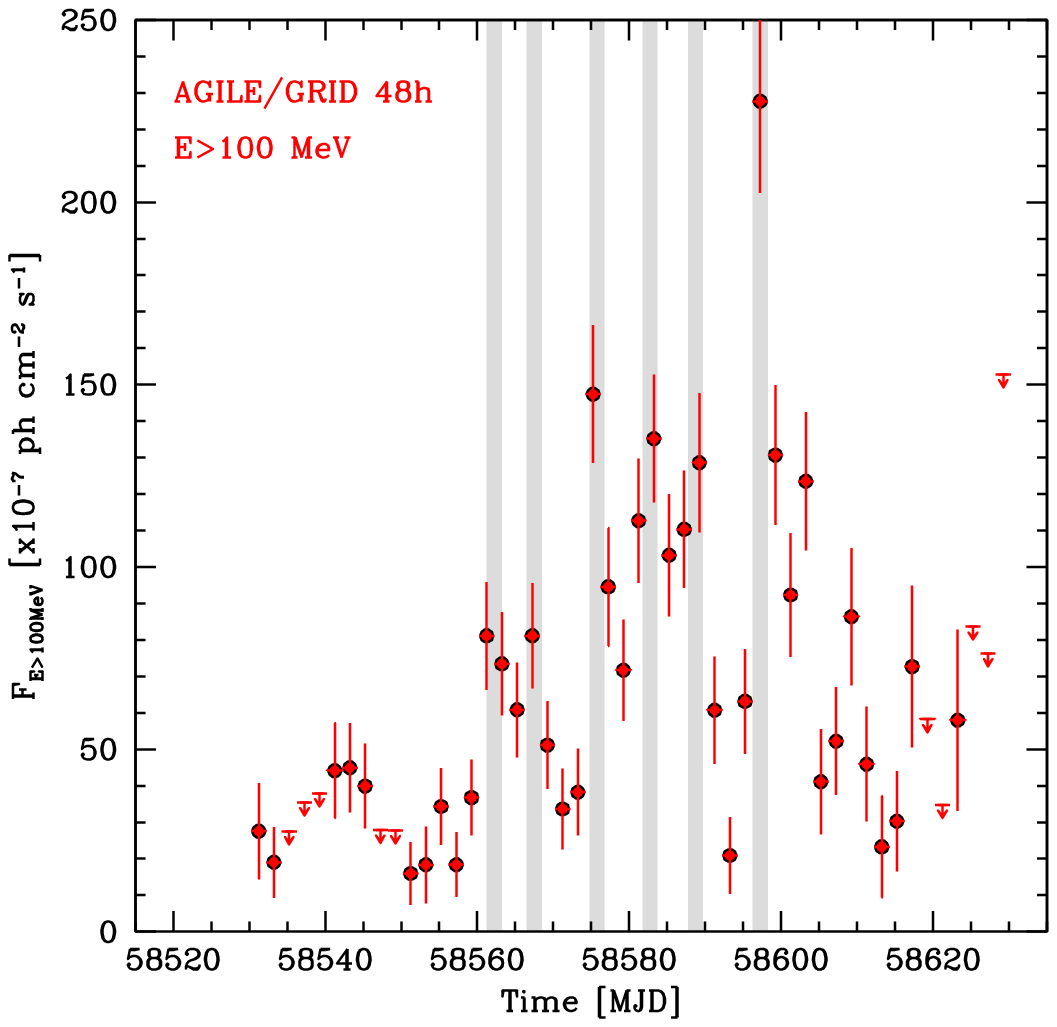}
      \caption{{One of the 8 statistically independent \grid{} 48h-bin light curves (E > 100 MeV) analyzed in our study, showing the maximum peak flux at MJD 58597.25 $\pm$ 1.0. Grey vertical bands correspond to the six high activity levels reported in Table~\ref{TAB:agileFlusGamma}}}
         \label{Fig:AGILE.LC}
   \end{figure}

We analyzed the \gray{} data above 100\,MeV between {February 16 and May 29, 2019 (MJD: 58530 - 58632). We analyzed 8 statistically independent light curves with a  48-hour time bin}, with a MSLA approach by calculating the flux at the nominal position of the blazar. {Each light curve is shifted by 6h with respect to the previous one, in order to better describe the time evolution of the \gray{} emission and preserving the photon statistics of a single 48h bin.}  In the multi-source analysis, we took into account the emission of the nearby sources within a radius of analysis of $10^{\circ}$. Position and fluxes of the field sources have been kept fixed at the values of the Second AGILE Catalog \citep{2019A&A...627A..13B}. {The parameter quantifying the Galactic diffuse emission has been kept fixed at a standard value for an extra-galactic source. The parameter related to the isotropic diffuse emission has been kept free to vary.
Figure~\ref{Fig:AGILE.LC} shows one of the eight independent light curves analyzed in our study, namely the one whose temporal bins are in agreement with the {\it Fermi}-LAT ones.}
%
%_____________________________________________________________
%_____________________________________________________________

%
%_____________________________________________________________
   \begin{figure}
   \centering
   \includegraphics[angle=0,width=9cm]{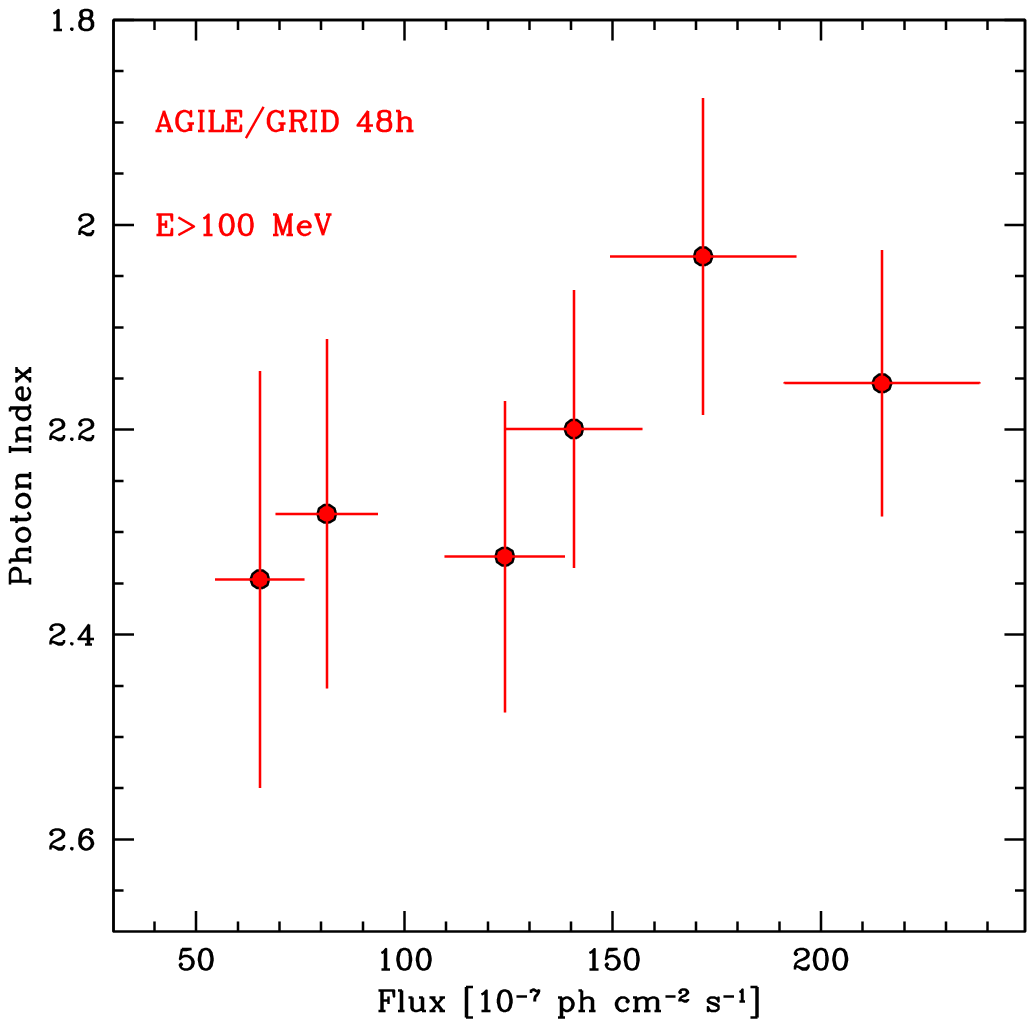}
      \caption{\grid{} photon index versus the $E>100$\,MeV flux. {See also Table~\ref{TAB:agileFlusGamma}.}}
         \label{Fig:AGILE.FvsG}
   \end{figure}
%_____________________________________________________________
%

%_____________________________________________________________________
% AGILE data
%----------------------------------------------------------------------
\begin{table}
\caption{\grid{} fluxes and photon indices at different relevant epochs.}     
\label{TAB:agileFlusGamma}      
\centering    
\footnotesize                
\begin{tabular}{lll}        
\hline                
Time interval       & F(E>100\,MeV) & Photon index \\
(MJD)      & (\phcmsec)                      & \\
\hline\\
58561.25 - 58563.25   &    $(8.1 \pm 1.2)\times 10^{-6}$  &  $2.28 \pm 0.17$ \\
58566.50 - 58568.50   &    $(6.5 \pm 1.1)\times 10^{-6}$  &  $2.35 \pm 0.20$ \\
58574.75 - 58576.75   &    $(1.2 \pm 0.1)\times 10^{-5}$  &  $2.32 \pm 0.15$ \\
58581.75 - 58583.75   &    $(1.4 \pm 0.2)\times 10^{-5}$  &  $2.20 \pm 0.14$ \\
58587.75 - 58589.75   &    $(1.7 \pm 0.2)\times 10^{-5}$  &  $2.03 \pm 0.16$ \\
58596.25 - 58598.25   &    $(2.2 \pm 0.2)\times 10^{-5}$  &  $2.16 \pm 0.13$ \\
\hline 
\end{tabular}
\end{table}
%--------------------------------------------------------------------
%____________________________________________________________________
In Fig.~\ref{Fig:AGILE.FvsG} and Table~\ref{TAB:agileFlusGamma}, we present the \grid{} photon index versus the $E>100$\,MeV flux, related to the main relative \gray{} peaks, emerging from the overall time evolution of the emission from the blazar (8 time-shifted light curves). The photon indices have been calculated with a binned analysis in the energy band 100 MeV - 3 GeV. Each flux has been calculated by keeping the correspondent spectral index fixed. No particular conclusions can be drawn on the correlation between the \gray{} flux and the photon index because of the too small number of data points (Spearman's coefficient $\rho=-0.89$, p$\simeq0.02$).

%%%%%%%%%%%%%%%%%%%%%%%%%%%%%%%%
\subsection{{\it Fermi}-LAT data}\label{FERMIdata}
%%%%%%%%%%%%%%%%%%%%%%%%%%%%%%%%

\begin{figure}
   \centering
   \includegraphics[angle=0,width=9cm]{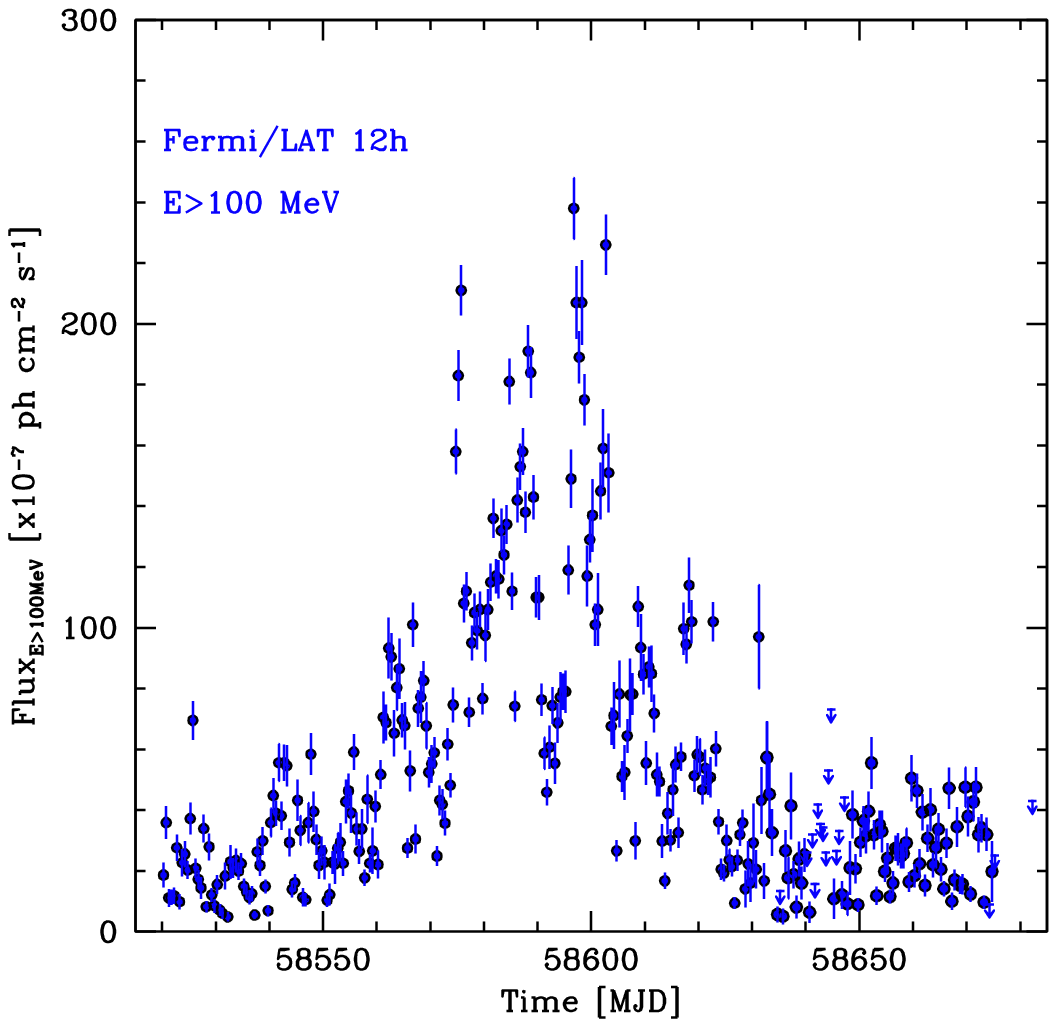}
      \caption{{\it Fermi}-LAT 12h-bin light curve obtained using likelihood standard analysis (E > 100 MeV) from 2019 February 06 to 2019 May 29.}
         \label{Fig:Fermi_12h_300MeV.LC}
   \end{figure}

\begin{figure}
   \centering
   \includegraphics[angle=0,width=9cm]{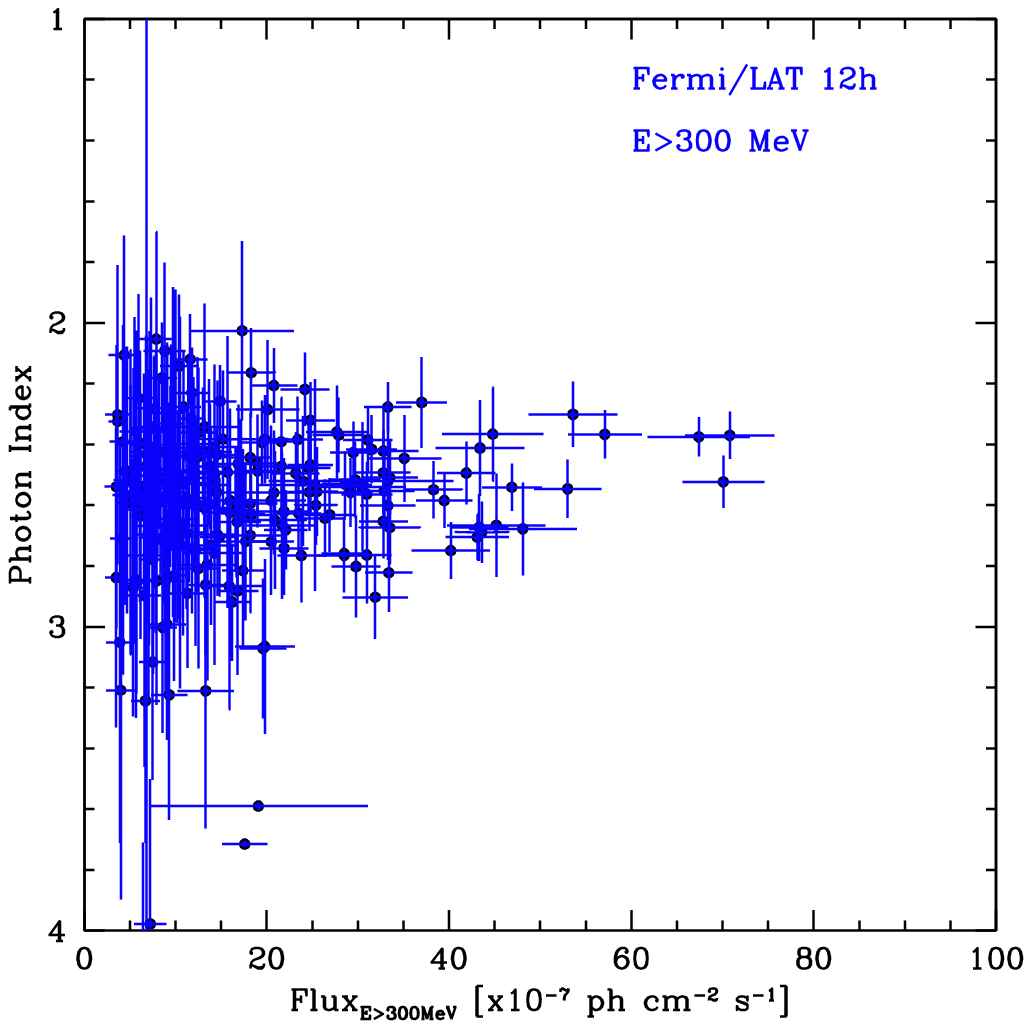}
      \caption{{\it Fermi}-LAT 12h-bin (E > 300 MeV) photon index versus 12h-bin flux from 2019 October 14 to 2019 May 20.}
         \label{Fig:Fermi_12h_flux_phind.LC}
   \end{figure}

\noindent
We analyzed the {\it Fermi}-LAT~\citep{2009ApJ...697.1071A} data using the standard tools provided with the \verb+ScienceTools+ version \verb+v11r05p02+, and the instrument response functions \verb+P8R3_SOURCE_V2+ to produce light curves and spectra. We selected events within a region of 20$^\circ$ around the source nominal position, with reconstructed energy in the 0.1-300\,GeV range. We filtered out $\gamma$-rays with zenith angles larger than 90$^\circ$ to reduce Earth limb $\gamma$-rays. We used the unbinned likelihood procedure to extract fluxes in energy and time bins. We modelled background using standard templates for isotropic and galactic diffuse background, and we included pointlike and diffuse sources from the fourth {\it Fermi}-LAT catalog \citep{2020ApJS..247...33A} inside the region of interest. Figure~\ref{Fig:Fermi_12h_300MeV.LC} and Figure~\ref{Fig:Fermi_12h_flux_phind.LC} show the {\it Fermi}-LAT 12h-bin light curve obtained using likelihood standard analysis (E > 100\,MeV)  and the 12h-bin (E > 300\,MeV) photon index versus 12h-bin flux, respectively.
We note that the (E > 300\,MeV) energy range used for the calculation of the spectral index is a conservative choice. The {\it Fermi}-LAT light-curves were calculated for both E > 100\,MeV and E > 300\,MeV, the former to allow a proper comparison with the \grid{} one.

%_____________________________________________________________
%_____________________________________________________________

%_____________________________________________________________
%_____________________________________________________________

%
%______________________________________________________________

%%%%%%%%%%%%%%%%%%%%%%%%%%%%%%%%
\section{X--ray Observations}\label{SEC:obs:Xrays}
%%%%%%%%%%%%%%%%%%%%%%%%%%%%%%%%
%
Table~\ref{TAB:all_Xray_Obs} summarizes all the X-ray/hard X-ray observations, including INTEGRAL, {\it Swift}-XRT, {\it Chandra}, and {\it NuSTAR}.
%
%
%
%_____________________________________________________________________
% All X observation log
%----------------------------------------------------------------------
\begin{table*}
\caption{Log of X-ray observations. MJDs are rounded for sake of simplicity. Actual MJD values can be derived from the ID column.}     
\label{TAB:all_Xray_Obs}      
\centering        
\begin{tabular}{lccc}        
\hline             
Date  & Satellite & ID & duration\\
(MJD)& & & (ks) \\
\hline
58545 & Swift-XRT &  00038422035 & 2.0 \\
58548 & Swift-XRT &  00038422036 & 2.0 \\
58550 & {\it NuSTAR} &  804606280002 & 40 \\
58551 & Swift-XRT &  00038422037 & 1.9 \\
58554 & Swift-XRT &  00038422038 & 2.0 \\
58558 & Swift-XRT &  00038422039 & 1.9 \\
58560 & Swift-XRT &  00038422040 & 1.7 \\
58563 & Swift-XRT &  00038422041 & 2.1 \\
58566 & Swift-XRT &  00038422042 & 2.0 \\
58567 & Swift-XRT &  00038422044 & 1.8 \\
58568 & Swift-XRT &  00038422045 & 2.0 \\
58572 & Swift-XRT &  00038422047 & 1.8 \\
58576 & Swift-XRT &  00038422049 & 0.8 \\
58576 & INTEGRAL (17 SCW) & 16700030001 &  --   \\
58577 & INTEGRAL (15 SCW)& 16700030001 & -- \\
58578 & Swift-XRT &  00038422050 & 1.6 \\
58578 & INTEGRAL (9 SCW)&  16700030001 & -- \\
58581 & Swift-XRT &  00038422051 & 2.0 \\
58583 & Swift-XRT &  00038422053 & 1.6 \\
58584 & Swift-XRT &  00038422054 & 2.2 \\
58585 & Swift-XRT &  00038422055 & 1.6 \\
58587 & Swift-XRT &  00038422056 & 2.0 \\
58590 & Swift-XRT &  00038422057 & 1.9 \\
58592 & Chandra & 22197 &  15 \\
58592 & INTEGRAL (6 SCW)&  16200150003 & -- \\
58593 & Swift-XRT &  00038422058 & 1.9 \\
58600 & Swift-XRT &  00038422059 & 0.01 \\
58602 & Swift-XRT &  00038422060 & 1.5 \\
58602 & Swift-XRT &  00038422061 & 0.4 \\
58606 & Swift-XRT &  00038422062 & 2.3 \\
58608 & INTEGRAL (1 SCW)&  16200150003 & -- \\
56609 & Swift-XRT &  00038422063 & 2.0 \\
58610 & Chandra & 22198 &  20 \\
58615 & Swift-XRT &  00038422064 & 2.6 \\
58618 & Swift-XRT &  00038422065 & 2.9 \\
58621 & Swift-XRT &  00038422066 & 2.3 \\
58627 & Swift-XRT &  00038422067 & 2.9 \\
58627 & Chandra & 22199 &  25 \\
58630 & Swift-XRT &  00038422068 & 2.6 \\
58633 & Swift-XRT &  00038422069 & 1.4 \\
58636 & Swift-XRT &  00038422070 & 3.3 \\
58639 & Swift-XRT &  00038422071 & 0.2 \\
58642 & Swift-XRT &  00038422072 & 2.7 \\
58645 & Swift-XRT &  00038422073 & 2.2 \\
58649 & Chandra & 22239 &  10 \\
58650 & Chandra & 22240 &  10 \\
58654 & Swift-XRT &  00038422074 & 1.5 \\
58658 & Swift-XRT &  00038422075 & 0.3 \\
58660 & Swift-XRT &  00038422076 & 1.1 \\
58665 & Swift-XRT &  00038422077 & 1.1 \\
\hline
\end{tabular}
\end{table*}
%--------------------------------------------------------------------
%____________________________________________________________________
%
%

%%%%%%%%%%%%%%%%%%%%%%%%%%%%%%%%
\subsection{INTEGRAL data}\label{INTEGRALdata}
%%%%%%%%%%%%%%%%%%%%%%%%%%%%%%%%
The INTEGRAL~\citep{1994ApJS...92..327W} data set consists of a 180\,ks public target of opportunity (ToO) observations performed from 03-April until 05-April 2019 plus 13\, ks of public General Program, in which the source was in  partially coded field of view, performed on 19-April-2019. The INTEGRAL data reduction of  the low energy detector, ISGRI of \gray{} telescope IBIS~\citep{2003A&A...411L.131U} was performed using  the standard Offline Scientific Analysis~\citep[OSA,][]{2003A&A...411L.223G} version 10.2 and the latest response matrices available. The source has been detected in the total ISGRI mosaic image (193 ks) at 7.1 and 7.2 sigma-level in the 15-30 keV and 30-200 keV energy ranges, respectively. The 20-50\,keV ISGRI flux is 0.47$\pm$0.07\,counts\,s$^{-1}$ (2.5$\times$ $10^{-11}$ erg cm$^{-2}$ s$^{-1}$ or 4.4\,mCrab). The 50-150\,keV ISGRI flux is 0.55$\pm$0.07\,counts\,s$^{-1}$ (7.9$\times$ $10^{-11}$ erg cm$^{-2}$ s$^{-1}$ or 10\,mCrab).
The ISGRI 20-50\,keV flux increased during the ToO observation by about 17$\%$ with respect to the averaged ISGRI flux reported in \cite{2011ApJ...736L..30D}, while the 40-100\,keV flux increment was of about 51$\%$, which is consistent with a hardening of the source spectrum.
We extracted the \pks{} ISGRI averaged spectrum (13-200\,keV) using both the standard OSA spectral extraction and the {alternative procedure to extract a faint source spectrum (see OSA user manual for details\footnote{\href{https://www.isdc.unige.ch/integral/download/osa/doc/11.1/osa\_um\_ibis.pdf}{https://www.isdc.unige.ch/integral/download/osa/doc/11.1/\\
osa\_um\_ibis.pdf}})}. The resulting spectra were consistent. The best fit model consists in a simple power law with the photon index $\Gamma$=1.0$\pm$0.3 and normalization $N_{\rm 0}=6.6^{+13}_{-2.1}\times10^{-4}$ {ph\,keV$^{-1}$\,cm$^{-2}$\,s$^{-1}$ at 1\,keV} ($\chi^{2}_{\rm red}$=1.0; 4 d.o.f.).
%

%%%%%%%%%%%%%%%%%%%%%%%%%%%%%%%%
\subsection{{\it Swift}-XRT data}\label{XRTdata}
%%%%%%%%%%%%%%%%%%%%%%%%%%%%%%%%

The {\it Neil Gehrels Swift} Observatory \citep[{\it Swift} hereafter, ][]{2004ApJ...611.1005G} data (Target ID 38422) were collected by activating two dedicated ToO observations triggered as a follow-up to AGILE detections. The X--ray Telescope \citep[XRT,][on-board {\it Swift}]{2005SSRv..120..165B} events files were processed using the \verb+XRTDAS+ software package (v.3.6.0) developed at SSDC and distributed by the High Energy Astrophysics Science Archive Research Center (HEASARC) within the HEASoft package.  Calibrated and cleaned event files were produced using the calibration files in the {\it Swift}-XRT {\tt CALDB 0(20200724)} and standard filtering criteria with the {\tt xrtpipeline} task. We used the {\tt xrtproducts} task included in the {\tt XRTDAS package} to extract the {\it Swift}-XRT source and background spectra using the appropriate response and ancillary files. We extracted spectra and light curves using circular apertures of radius 30$\asec$, centered on the peak of the emission in the 0.3-10\,keV data. Background spectra were extracted using source--free annular regions of the 80/120\, pixel inner/outer radius.

\begin{figure}
   \centering
      \includegraphics[angle=0,width=9cm]{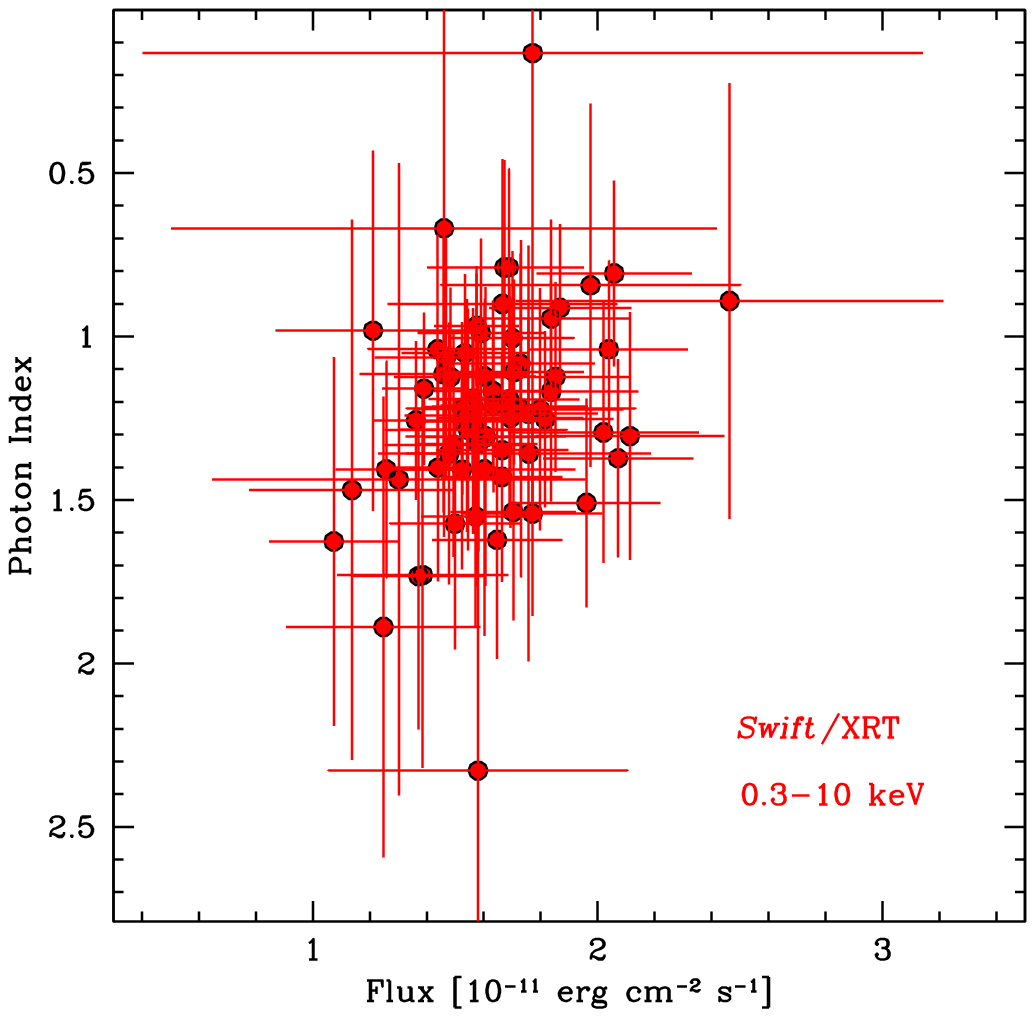}
      \caption{{\it Swift}/XRT photon index s a function of the 0.3-10\,keV observed flux.}
         \label{Fig:XRT_FGAMMA}
   \end{figure}

Figure~\ref{Fig:XRT_FGAMMA} shows the {\it Swift}/XRT photon index as function of the 0.3-10\,keV observed flux. A possible harder-when-brighter trend is present (Spearman's coefficient $\rho=-0.31$, p<0.009).

%_____________________________________________________________
%_____________________________________________________________

%%%%%%%%%%%%%%%%%%%%%%%%%%%%%%%%
\subsection{{\it Chandra} data}\label{CHANDRAdata}
%%%%%%%%%%%%%%%%%%%%%%%%%%%%%%%%
%
{\it Chandra}~\citep{1994ApJS...92..327W} pointed at \pks{} on discretionary director time (DDT) observations. Five observations were acquired, three in April and May 2019 and two in June 2019. Table \ref{chobs.tab} shows the log of the {\it Chandra} observations.
We reprocessed the {\it Chandra} data using the ``{\it Chandra} Interactive Analysis of Observations'' ({\verb+CIAO+}) package. For each observation, we created the clean level-2 event file using the {\tt chandra\_repro} script. \citet{2005A&A...438..121D}  reported that {\it Chandra} can resolve the two lensed images of \pks{} at an angular distance of the order of 1\arcsec. In order to search for possible spectral differences of the two hotspots (see below), we first align the WCS grid of all observations to the same reference coordinate. We follow the standard \verb+CIAO+ thread for absolute astrometric correction. For all observations, we use the {\tt wcs\_match} tool to compute the offset between the observed centroid and the reference coordinates. Hence, this can be input in {\tt wcs\_update} script to update the WCS grid of all the event files to match the reference coordinates {(i.e. the coordinates of the source as observed in the first observation).}

\begin{figure}
   \centering
      \includegraphics[angle=0,width=9cm]{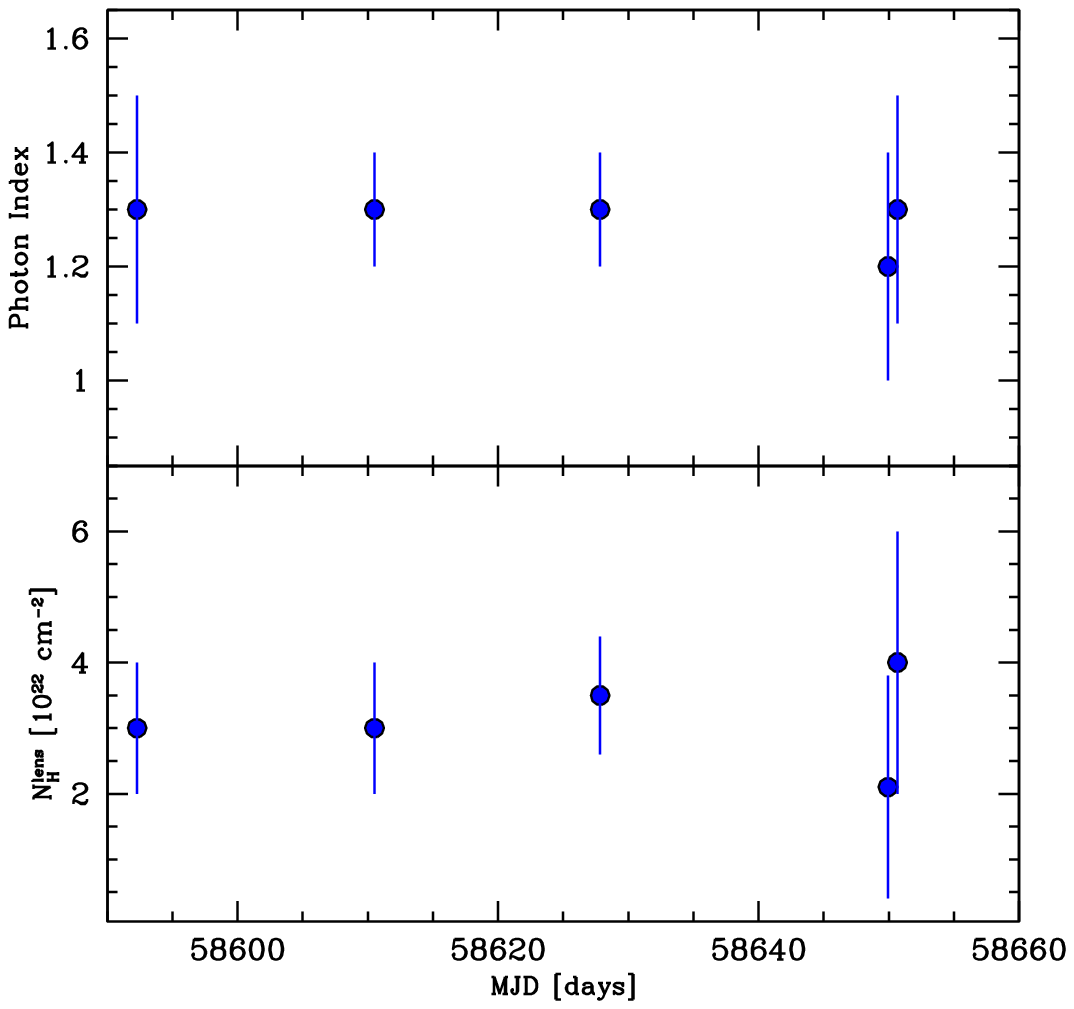}
      \caption{Time evolution of the parameters of the {\it Chandra} spectra.
      Upper panel: Photon index.
      Lower panel: Absorbing column density at the redshift of the
      lensing galaxy.}
         \label{chspecpar.fig}
   \end{figure}

For all observations, we extracted the spectrum of \pks{} from a circular region with a radius of 4\arcsec. We took the background from an annular region, centered on the source, and with inner and outer radii of 20\arcsec{} and 25\arcsec, respectively. We fitted the spectra using {\tt XSpec v12}. It was already reported that \pks{} displays a highly absorbed spectrum, with the absorption arising in the intervening lensing galaxy at z=0.89. Thus, we fitted all the spectra with a model that included the Galactic absorption \citep[$N_{\rm H}^{\rm gal}=2.19 \times 10^{21}$ \colc,][]{1990ARA&A..28..215D}, the absorption at the redshift of the lensing galaxy, and a power law continuum. The free parameters of the fits are the intervening column density, $N_{\rm H}^{\rm lens}$, the photon index,  and the normalization of the power law. Figure~\ref{chspecpar.fig} shows the time evolution of the free parameters of our fits, observation by observation, while the observed fluxes measured from our spectral fitting are reported in Table \ref{chobs.tab}. In all the {\it Chandra} observations, the global X-ray spectrum of \pks{} remained substantially  stable, both in flux and in spectral shape.
%
%_____________________________________________________________
  \begin{figure}
   \centering\includegraphics[angle=0,width=9cm]{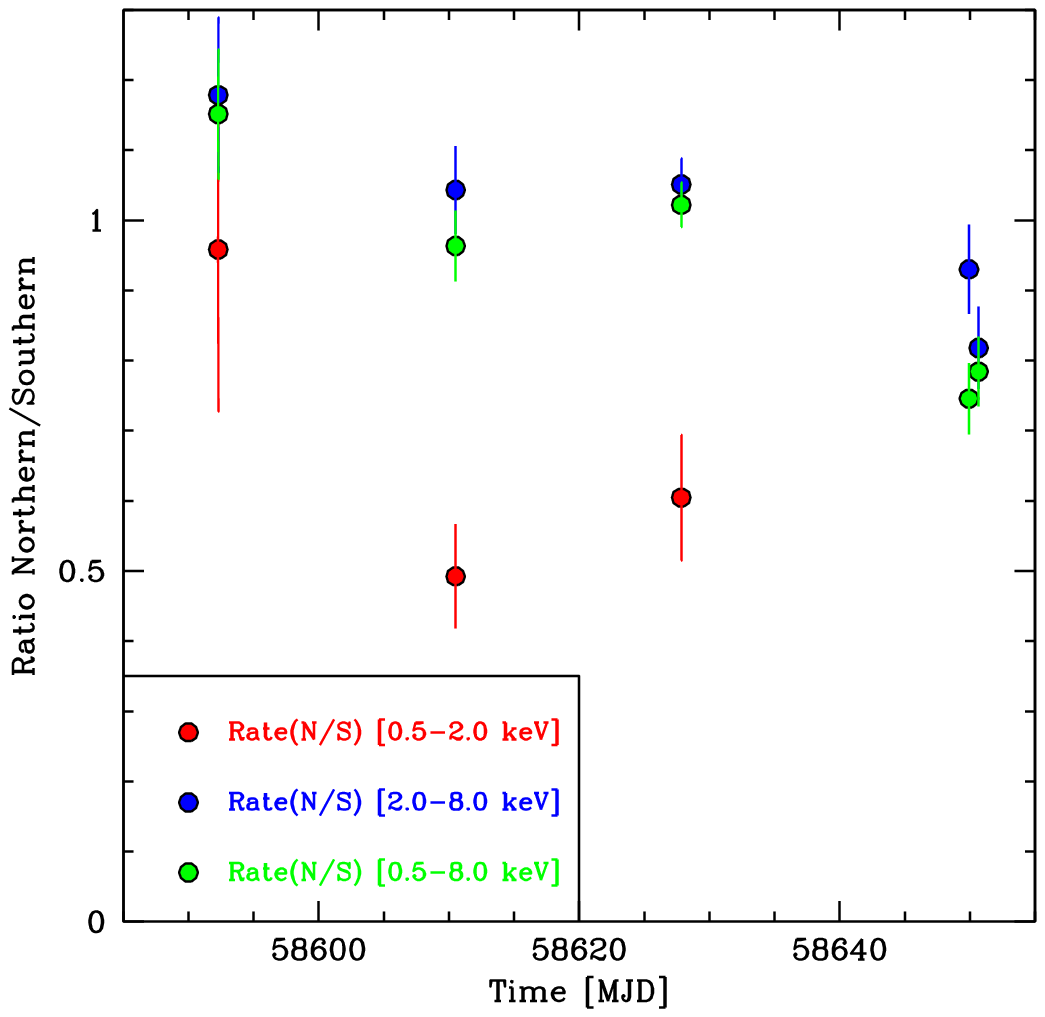}
      \caption{Ratios between the count-rate measured in the north and south hotspot of PKS~1830$-$211 as a function of time in the 0.5-2.0 (red), 2.0-8.0 (blue) and 0.5-8.0\,keV (green) energy range.}
         \label{Fig:CHANDRA_HOTSPOTS}
   \end{figure}
%_____________________________________________________________
%

%
%_____________________________________________________________________
% Chandra observation log
%----------------------------------------------------------------------
\begin{table}
\caption{{\it Chandra} observation log}     
\label{chobs.tab}      
\centering                    
\begin{tabular}{llll}        
\hline                 
Obs. ID & Date & Duration$^{a}$ & Observed Flux$^{b}$\\
& dd/mm/yyyy & & $F_{\rm 0.5-2.0 keV}$ \quad $F_{\rm 2.0-10.0 keV}$ \\
\hline
22197 & 19/04/2019 & 15 & $0.11\pm0.03$ \quad $1.3\pm0.4$\\
22198 & 07/05/2019 & 20 & $0.15\pm0.03$ \quad $1.6 \pm 0.3$ \\
22199 & 24/05/2019 & 25 & $0.11\pm0.02$ \quad $1.3 \pm 0.2$ \\
22239 & 15/06/2019 & 10 & $0.12\pm0.03$ \quad $1.4 \pm 0.3$ \\
22240 & 16/06/2019 & 10 & $0.12\pm0.03$ \quad $1.4 \pm 0.3$ \\
\hline 
\multicolumn{4}{l}{$^a$ Total duration of the observation in ksec.}\\
\multicolumn{4}{l}{$^b$ Observed flux in the quoted bands. Units of $10^{-11}$ \ergsc.}\\
\end{tabular}
\end{table}
%--------------------------------------------------------------------
%____________________________________________________________________

%_____________________________________________________________
%_____________________________________________________________

%%%%%%%%%%%%%%%%%%%%%%%%%%%%%%%%
\subsection{{\it NuSTAR} data}\label{NUSTARdata}
%%%%%%%%%%%%%%%%%%%%%%%%%%%%%%%%

{\it NuSTAR}~\citep{2013ApJ...770..103H} consists of two focal plane modules, FPMA and FPMB, is sensitive at 3--78.4~keV and has a field-of-view of 10\amin{} at 10~keV~\citep{2013ApJ...770..103H}. {\it NuSTAR} has a 18\arcsec{} FWHM with a half--power diameter of 58\arcsec. We analyzed the {\it NuSTAR} observation of \pks{} performed in March 2019. The observation log is given in Table~\ref{tablog}.

\begin{table*}
\footnotesize
\caption{{\it NuSTAR} observation log.}
\centering
\begin{tabular}{lcccccccc}
\hline
Observation ID$^a$ & RA\_PNT$^b$ & DEC\_PNT$^c$ &  Exposure$^d$ & Start Date$^e$ & rate$^f$& background$^g$ \\ 
 & (deg.) & (deg.) & (ksec) &  &  (cts/s) & \\ 
\hline
 80460628002 &   278.4356 &  -21.0336  &  41.4  & 2019-03-08T20:21:09 & 0.458$\pm$0.004 & $\sim$6\%\\ [1ex]
\hline 
\end{tabular}

Notes: $^a$Observation identification number; $^b$Right Ascension of the pointing; $^c$Declination of the pointing; $^d$ total net exposure time; $^e$start date and time of the observation; $^f$mean value of the net count rate in the circular source extraction region with 90\arcsec radius in the energy range 3--78.4 keV; $^g$ background percentage in the circular source extraction region with 90\arcsec radius and in the energy range 3--78.4 keV.
\label{tablog}
\end{table*}

The raw events files were processed using the {\it NuSTAR} Data Analysis Software package v. 2.0.0. {\tt NuSTARDAS}\footnote{\href{http://heasarc.gsfc.nasa.gov/docs/nustar/analysis/nustar\_swguide.pdf}{http://heasarc.gsfc.nasa.gov/docs/nustar/analysis/nustar\_swguide.pdf}}. Calibrated and cleaned event files were produced using the calibration files in the {\it NuSTAR} {\tt CALDB} (20200813) and standard filtering criteria with the {\tt nupipeline} task.  We used {\tt nuproducts} task included in the {\tt NuSTARDAS} package to extract the {\it NuSTAR} source and background spectra using the appropriate response and ancillary files. We extracted spectra and light curves in each focal plane module (FPMA and FPMB) using circular apertures of radius 90\arcsec, corresponding to $\sim 90\%$ of the encircled energy, centered on the peak of the emission in the 3--78.4~keV data. Background spectra were extracted using source--free regions on the same detector as the source. As shown in Table~\ref{tablog}, the background count rates are a small fraction ($\sim 6\%$) of the source count rates.  The spectra were binned to have at least 30 counts per bin.

%%%%%%%%%%%%%%%%%%%%%%%%%%%%%%%%
\section{IR-Optical-UV Observations}\label{SEC:obs:iropt}
%%%%%%%%%%%%%%%%%%%%%%%%%%%%%%%%
%

%%%%%%%%%%%%%%%%%%%%%%%%%%%%%%%%
\subsection{{\it Swift}/UVOT data}\label{UVOTdata}
\begin{figure}
   \centering
      \includegraphics[angle=0,width=8cm]{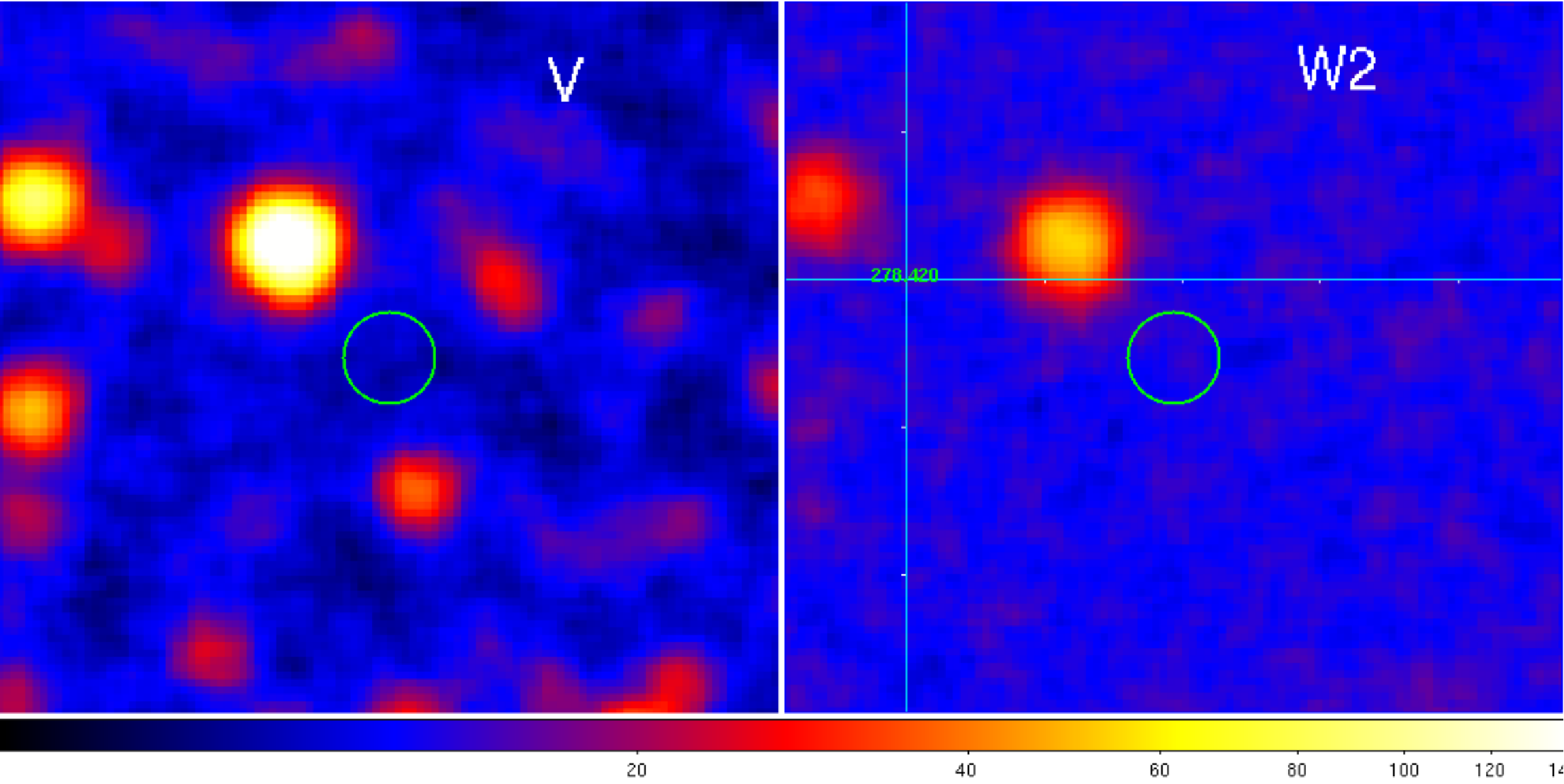}
      \caption{UVOT image for the V (left) and W2 filters. The non-standard aperture of size 2.2$\arcsec$ is shown in cyan color.}
         \label{Fig:UVOT_VW2}
   \end{figure}

%%%%%%%%%%%%%%%%%%%%%%%%%%%%%%%%
%
The {\it Swift}/UVOT data were accumulated during the whole observing campaign, in order to establish the most reliable upper limits in the $v$, $b$, $u$, $w1$, $m2$, and $w2$ filters. 
Data were processed with {\tt HEAsoft v6.23} and {\tt CALDB (20201026)}. 
Due to the crowded field, particularly in optical bands, some nearby stars could contaminate an aperture of standard radius (see Figure~\ref{Fig:UVOT_VW2}). To extract source counts we used a non-standard aperture of radius equal to the PSF FWHM,  2.2$\arcsec$, and  three uncontaminated circular regions for the background extraction. We extracted source fluxes in each filter image available in each observation and on the sum of all the available images during the 2019 flaring campaign in each filter. No detections were obtained in any band, so we could not confirm the detection reported in~\cite{2021ApJ...915...26A}.
Observed magnitudes were converted into {dereddened} fluxes according to the {\tt CALDB} conversion factors~\citep{brev11} {and a mean Galactic extinction law~\citep{1999PASP..111...63F} and E(B-V) value of 0.397~\citep{2011ApJ...737..103S}. Flux ULs on the summed images are reported in Table~\ref{Tab:uvot}.}

%_____________________________________________________________________
% UVOT data
%_____________
%_____________
%----------------------------------------------------------------------
\begin{table}
\caption{{\it Swift/}UVOT data.}     
\label{Tab:uvot}      
\centering                    
\begin{tabular}{ll}        
\hline                 
Filter & Upper limits (dereddened)\\
         & (\ergsc)\\
\hline
$v$     & $< 1.80\times 10^{-13}$\\
$b$     &  $< 8.96\times 10^{-14}$\\
$u$     &   $< 2.42\times 10^{-14}$\\
$w1$   &   $< 1.74\times 10^{-14}$\\
$m2$    &  $<  1.29\times 10^{-14}$\\
$w2$    &  $<  1.45\times 10^{-14}$\\
\hline 
\end{tabular}
\end{table}
%--------------------------------------------------------------------
%____________________________________________________________________

%%%%%%%%%%%%%%%%%%%%%%%%%%%%%%%%
\subsection{REM data}\label{REMdata}
%%%%%%%%%%%%%%%%%%%%%%%%%%%%%%%%
%
%
{The infrared J, H, and K data have been obtained with the REMIR camera \citep{2003SPIE.4841..627V}, at the focal plane of the REM telescope \citep{2004SPIE.5492.1590Z}, in the ESO La Silla (Chile) observatory. Observations were carried out in 23 nights, in March and April 2019, of which 12 were photometric. The NIR frames have been reduced with the semi-automatic pipeline {\tt Riace} (Giannini et al. {\it in preparation}), which performs the frames stack (to produce the science and sky frames), then uses the 2MASS stars in the field to calibrate the aperture photometry of the science target.
}
Figure~\ref{Fig:MWL.LC} panel (b) shows the REM light-curves for the H and K filters. Data are reported in Table~\ref{Tab:REM}.

%_____________________________________________________________________
% UVOT data
%----------------------------------------------------------------------
\begin{table}
\caption{REM IR data.}     
\label{Tab:REM}      
\centering    
\footnotesize                
\begin{tabular}{lll}        
\hline                 
Time        & H-filter  & K-filter     \\
        & (dereddened) &  (dereddened) \\
(MJD)      & (\ergsc)                      & (\ergsc) \\
\hline
{58565.0 }   &            {$1.64\times10^{-13} \pm  6.19\times10^{-14}$}  &  -- \\
58566.0    &            $1.98\times10^{-13} \pm   7.45\times10^{-14}$  &   $2.54\times10^{-13} \pm   9.57\times10^{-14}  $ \\
58571.0    &            $2.17\times10^{-13} \pm   1.03\times10^{-13}$  &   $2.32\times10^{-13} \pm   8.73\times10^{-14}  $ \\
58573.0    &            $2.38\times10^{-13} \pm   6.65\times10^{-14}$  &   $2.79\times10^{-13} \pm   1.05\times10^{-13}  $ \\
58574.0    &             $2.17\times10^{-13} \pm   8.16\times10^{-14}$  &   $2.54\times10^{-13} \pm   7.11\times10^{-14}  $ \\
58577.0    &             --  &   $3.05\times10^{-13} \pm  5.66\times10^{-14}  $ \\
58578.0    &             $2.17\times10^{-13} \pm   6.06\times10^{-14}$  &   $2.79\times10^{-13} \pm   5.16\times10^{-14}  $ \\
58580.0    &             $2.38\times10^{-13} \pm   6.65\times10^{-14}$  &   $2.54\times10^{-13} \pm   4.71\times10^{-14}  $ \\
58584.0    &             $1.50\times10^{-13} \pm   5.65\times10^{-14}$  &   $1.46\times10^{-13} \pm   4.09\times10^{-14}  $ \\
58585.0    &             $1.80\times10^{-13} \pm   5.04\times10^{-14}$  &   $1.93\times10^{-13} \pm   5.39\times10^{-14}  $ \\
58586.0    &             --  &   $1.93\times10^{-13} \pm   5.39\times10^{-14}  $ \\
\hline 
\end{tabular}
\end{table}
%--------------------------------------------------------------------
%____________________________________________________________________

%%%%%%%%%%%%%%%%%%%%%%%%%%%%%%%%
\section{Radio Observations}\label{SEC:obs:radio}
%%%%%%%%%%%%%%%%%%%%%%%%%%%%%%%%

%%%%%%%%%%%%%%%%%%%%%%%%%%%%%%%%
\subsection{SRT/Medicina data}\label{SRTMEDdata}
%%%%%%%%%%%%%%%%%%%%%%%%%%%%%%%%
%
A radio observation campaign was undertaken with two INAF radio telescopes: the Sardinia Radio Telescope\footnote{\href{http://www.srt.inaf.it/}{http://www.srt.inaf.it/}} (SRT) and the Medicina radio telescope\footnote{\href{http://www.med.ira.inaf.it/}{http://www.med.ira.inaf.it/}}. The radio follow-up started on MJD 58590.2 and ended on MJD 58662.9. 18 dual-frequency observing sessions were performed: 8 with Medicina at 8.3 and 25.4\,GHz using {\tt total power} back-end with 250 and 680\,MHz bandwidth, respectively, and 10 sessions with the SRT at 7 and 25.5\,GHz, where {\tt SARDARA} back-end \citep{2018JAI.....750004M} was used with a 1400\,MHz bandwidth. Single-dish radio mapping techniques were exploited to perform On The Fly maps of the source and a sample of the best-known radio astronomical flux calibrators (3C~286, 3C~295 and NGC~7027). 
The follow-up was undertaken in the context of two INAF ToO programs and one Director Discretionary Time proposal. Preliminary results were published in \cite{2019ATel12667....1I}. Radio imaging data analysis and calibration were performed using the techniques explained in \cite{2017MNRAS.471.2703E, 2017MNRAS.470.1329E} and \cite{2019MNRAS.482.3857L}, comparing counts of the Gaussian fit in the target image with calibrators images and cross-scans. Most of the radio data were of good quality. However, the low elevation of the source, bad weather (fog and rain) and strong radio frequency interference (RFI) affected the data scans (i.e. enhanced and variable error bars in flux densities). Figure~\ref{Fig:MWL.LC}, panel (a), shows the multi-frequency radio flux density light-curve (in Jy) obtained with SRT and Medicina radio telescopes in the MJD 58590.2-58662.9 period.
%

%_____________________________________________________________________
% SRT/MED data
%----------------------------------------------------------------------
\begin{table}
\caption{SRT and Medicina radio data.}     
\label{Tab:SRTMED}      
\centering    
\begin{tabular}{lll}        
\hline        
\multicolumn{3}{c}{SRT times and data}  \\   
Time        & 7\,GHz & 25.5\,GHz \\
(MJD)      & (Jy)                      & (Jy) \\
\hline
58596.253   &    $15.8 \pm 1.0$   &    $11.9 \pm 0.8$ \\
58599.193   &    $15.2 \pm 1.0$   &    $11.7 \pm 1.0$ \\
58601.226   &    $14.5 \pm 0.8$   &    $11.7 \pm 1.5$ \\
58604.208   &    $14.6 \pm 0.9$   &    $11.5 \pm 1.5$ \\ 
58607.184   &    $14.0 \pm 0.9$   &    $11.2 \pm 1.0$ \\
58610.191   &    $14.6 \pm 0.8$   &    $11.3 \pm 0.8$ \\
58613.190   &    $15.2 \pm 0.8$   &    $11.9 \pm 1.3$ \\
58617.147   &    $14.9 \pm 1.0$   &    $11.6 \pm 1.0$ \\
58626.113   &    $15.4 \pm 1.5$   &    $11.5 \pm 0.5$ \\
58630.169   &    $15.7 \pm 1.0$   &    $10.5 \pm 1.0$ \\
\multicolumn{3}{c}{}  \\   
\multicolumn{3}{c}{Medicina times and data}  \\   
Time        & 8.3\,GHz & 25.5\,GHz \\
(MJD)      & (Jy)                      & (Jy) \\
\hline
58590.228   &    $16.0 \pm 1.0$   &    $9.8 \pm 0.9$ \\
58592.236   &    $17.0 \pm 1.0$   &    $11.8 \pm 0.5$ \\
58644.900   &    $16.0 \pm 1.0$   &    $10.2 \pm 0.5$ \\
58649.941   &    $17.0 \pm 0.5$   &    $10.7 \pm 0.5$ \\
58651.925   &    $17.0 \pm 0.7$   &    $10.7 \pm 0.9$ \\
58655.902   &    $16.5 \pm 0.4$   &    $9.7 \pm 0.6$ \\
58659.902   &    $17.1 \pm 0.4$   &    $9.8 \pm 0.6$ \\
58662.905   &    $17.6 \pm 0.5$   &    $10.7 \pm 1.0$ \\
\hline 
\end{tabular}
\end{table}
%--------------------------------------------------------------------
%____________________________________________________________________

%
%

%%%%%%%%%%%%%%%%%%%%%%%%%%%%%%%%
\subsection{OVRO data}\label{OVROdata}
%%%%%%%%%%%%%%%%%%%%%%%%%%%%%%%%
%
\noindent
The Owens Valley Radio Observatory (OVRO) 40-Meter Telescope uses off-axis dual-beam optics and a cryogenic receiver with 2~GHz equivalent noise bandwidth centered at 15~GHz. Gain fluctuations, atmospheric and ground contributions are removed with the double switching technique~\citep{1989ApJ...346..566R} where the observations are conducted in an ON-ON fashion such that one of the beams is always pointed on the source. The two beams were rapidly alternated using a Dicke switch until May 2014. In May 2014 a new pseudo-correlation receiver with a 180~degree phase switch replaced the old receiver. To compensate for gain drifts relative calibration is obtained with a temperature-stable noise diode. The primary flux density calibrator is 3C~286 with an assumed value of 3.44\,Jy~\citep{1977A&A....61...99B}, DR21 is used as secondary calibrator source. \citet{2011ApJS..194...29R} gives details about the observation procedure and data reduction. Figure~\ref{Fig:MWL.LC}, panel (a), shows the OVRO radio flux density light-curve (in Jy).

%%%%%%%%%%%%%%%%%%%%%%%%%%%%%%%%%%%%%%%%%%%%%%%%%%

% Don't change these lines
\bsp	% typesetting comment
\label{lastpage}
\end{document}